\documentclass[12pt,a4paper]{article}
\usepackage{ifthen} % for conditional statements
\newboolean{pdflatex}
\setboolean{pdflatex}{true} % False for eps figures 

\newboolean{articletitles}
\setboolean{articletitles}{true} % False removes titles in references

\newboolean{uprightparticles}
\setboolean{uprightparticles}{false} %True for upright particle symbols

\newboolean{inbibliography}
\setboolean{inbibliography}{false} %True once you enter the bibliography

\usepackage{microtype}
\usepackage{lineno}  % for line numbering during review
\usepackage{xspace} % To avoid problems with missing or double spaces after
\usepackage{caption} %these three command get the figure and table captions automatically small

\usepackage{graphicx}  % to include figures (can also use other packages)
\usepackage{color}
\usepackage{colortbl}
\graphicspath{{./figs/}} % Make Latex search fig subdir for figures

\usepackage{amsmath} % Adds a large collection of math symbols
\usepackage{amssymb}
\usepackage{amsfonts}
\usepackage{upgreek} % Adds in support for greek letters in roman typeset

\newcommand*\patchAmsMathEnvironmentForLineno[1]{%
\expandafter\let\csname old#1\expandafter\endcsname\csname #1\endcsname
\expandafter\let\csname oldend#1\expandafter\endcsname\csname
end#1\endcsname
 \renewenvironment{#1}%
   {\linenomath\csname old#1\endcsname}%
   {\csname oldend#1\endcsname\endlinenomath}%
}
\newcommand*\patchBothAmsMathEnvironmentsForLineno[1]{%
  \patchAmsMathEnvironmentForLineno{#1}%
  \patchAmsMathEnvironmentForLineno{#1*}%
}
\AtBeginDocument{%
\patchBothAmsMathEnvironmentsForLineno{equation}%
\patchBothAmsMathEnvironmentsForLineno{align}%
\patchBothAmsMathEnvironmentsForLineno{flalign}%
\patchBothAmsMathEnvironmentsForLineno{alignat}%
\patchBothAmsMathEnvironmentsForLineno{gather}%
\patchBothAmsMathEnvironmentsForLineno{multline}%
\patchBothAmsMathEnvironmentsForLineno{eqnarray}%
}

\usepackage{hyperref}    % Hyperlinks in references
\usepackage[all]{hypcap} % Internal hyperlinks to floats.

\usepackage{xspace} 
\usepackage{upgreek}

\def\lhcb {\mbox{LHCb}\xspace}

\def\MagUp {\mbox{\em Mag\kern -0.05em Up}\xspace}

\ifthenelse{\boolean{uprightparticles}}%
{

 \def\PDelta      {\ensuremath{\Delta}\xspace}                 
 \def\PXi      {\ensuremath{\Xi}\xspace}                 
 \def\PLambda      {\ensuremath{\Lambda}\xspace}                 
 \def\PSigma      {\ensuremath{\Sigma}\xspace}                 
 \def\POmega      {\ensuremath{\Omega}\xspace}                 
 \def\PUpsilon      {\ensuremath{\Upsilon}\xspace}                 
 
 \def\PB      {\ensuremath{\mathrm{B}}\xspace}                 
                  
 \def\PD      {\ensuremath{\mathrm{D}}\xspace}

 \def\PK      {\ensuremath{\mathrm{K}}\xspace}

 \def\PW      {\ensuremath{\mathrm{W}}\xspace}

 \def\PZ      {\ensuremath{\mathrm{Z}}\xspace}                 
                  
 \def\Pb      {\ensuremath{\mathrm{b}}\xspace}                 
 \def\Pc      {\ensuremath{\mathrm{c}}\xspace}

 \def\Pi      {\ensuremath{\mathrm{i}}\xspace}

}
{

 \mathchardef\PDelta="7101
 \mathchardef\PXi="7104
 \mathchardef\PLambda="7103
 \mathchardef\PSigma="7106
 \mathchardef\POmega="710A
 \mathchardef\PUpsilon="7107
                  
 \def\PB      {\ensuremath{B}\xspace}                 
                  
 \def\PD      {\ensuremath{D}\xspace}

 \def\PK      {\ensuremath{K}\xspace}

 \def\PW      {\ensuremath{W}\xspace}

 \def\PZ      {\ensuremath{Z}\xspace}                 
                  
 \def\Pb      {\ensuremath{b}\xspace}                 
 \def\Pc      {\ensuremath{c}\xspace}

 \def\Pi      {\ensuremath{i}\xspace}

}

\makeatletter
\ifcase \@ptsize \relax% 10pt
  \newcommand{\miniscule}{\@setfontsize\miniscule{4}{5}}% \tiny: 5/6
\or% 11pt
  \newcommand{\miniscule}{\@setfontsize\miniscule{5}{6}}% \tiny: 6/7
\or% 12pt
  \newcommand{\miniscule}{\@setfontsize\miniscule{5}{6}}% \tiny: 6/7
\fi
\makeatother

\DeclareRobustCommand{\optbar}[1]{\shortstack{{\miniscule (\rule[.5ex]{1.25em}{.18mm})}
  \\ [-.7ex] $#1$}}

\def\W      {{\ensuremath{\PW}}\xspace}
\def\Wp     {{\ensuremath{\PW^+}}\xspace}
\def\Wm     {{\ensuremath{\PW^-}}\xspace}
\def\Wpm    {{\ensuremath{\PW^\pm}}\xspace}
\def\Z      {{\ensuremath{\PZ}}\xspace}

\def\cquark    {{\ensuremath{\Pc}}\xspace}

\def\bquark    {{\ensuremath{\Pb}}\xspace}

  \def\Kbar    {{\kern 0.2em\overline{\kern -0.2em \PK}{}}\xspace}

\def\KorKbar    {\kern 0.18em\optbar{\kern -0.18em K}{}\xspace}

  \def\Dbar    {{\kern 0.2em\overline{\kern -0.2em \PD}{}}\xspace}

\def\DorDbar    {\kern 0.18em\optbar{\kern -0.18em D}{}\xspace}

\def\Bbar    {{\ensuremath{\kern 0.18em\overline{\kern -0.18em \PB}{}}}\xspace}

\def\BorBbar    {\kern 0.18em\optbar{\kern -0.18em B}{}\xspace}

  \def\Y#1S{\ensuremath{\PUpsilon{(#1S)}}\xspace}% no space before {...}!

\def\Lbar        {{\ensuremath{\kern 0.1em\overline{\kern -0.1em\PLambda}}}\xspace}
\def\LorLbar    {\kern 0.18em\optbar{\kern -0.18em \PLambda}{}\xspace}

\def\to                 {\ensuremath{\rightarrow}\xspace}

\newcommand{\as}{{\ensuremath{\alpha_s}}\xspace}

\def\AT#1     {\ensuremath{A_{\mathrm{T}}^{#1}}\xspace}           % 2

\def\C#1      {\ensuremath{\mathcal{C}_{#1}}\xspace}                       % 9
\def\Cp#1     {\ensuremath{\mathcal{C}_{#1}^{'}}\xspace}                    % 7
\def\Ceff#1   {\ensuremath{\mathcal{C}_{#1}^{\mathrm{(eff)}}}\xspace}        % 9  
\def\Cpeff#1  {\ensuremath{\mathcal{C}_{#1}^{'\mathrm{(eff)}}}\xspace}       % 7
\def\Ope#1    {\ensuremath{\mathcal{O}_{#1}}\xspace}                       % 2
\def\Opep#1   {\ensuremath{\mathcal{O}_{#1}^{'}}\xspace}                    % 7

\newcommand{\tev}{\ifthenelse{\boolean{inbibliography}}{\ensuremath{~T\kern -0.05em eV}\xspace}{\ensuremath{\mathrm{\,Te\kern -0.1em V}}}\xspace}
\newcommand{\gev}{\ensuremath{\mathrm{\,Ge\kern -0.1em V}}\xspace}
\newcommand{\mev}{\ensuremath{\mathrm{\,Me\kern -0.1em V}}\xspace}
\newcommand{\kev}{\ensuremath{\mathrm{\,ke\kern -0.1em V}}\xspace}
\newcommand{\ev}{\ensuremath{\mathrm{\,e\kern -0.1em V}}\xspace}
\newcommand{\gevc}{\ensuremath{{\mathrm{\,Ge\kern -0.1em V\!/}c}}\xspace}
\newcommand{\mevc}{\ensuremath{{\mathrm{\,Me\kern -0.1em V\!/}c}}\xspace}
\newcommand{\gevcc}{\ensuremath{{\mathrm{\,Ge\kern -0.1em V\!/}c^2}}\xspace}
\newcommand{\gevgevcccc}{\ensuremath{{\mathrm{\,Ge\kern -0.1em V^2\!/}c^4}}\xspace}
\newcommand{\mevcc}{\ensuremath{{\mathrm{\,Me\kern -0.1em V\!/}c^2}}\xspace}

\def\mm   {\ensuremath{\mathrm{ \,mm}}\xspace}

\def\mum  {\ensuremath{{\,\upmu\mathrm{m}}}\xspace}

\def\pb {\ensuremath{\mathrm{ \,pb}}\xspace}

\def\gsim{{~\raise.15em\hbox{$>$}\kern-.85em
          \lower.35em\hbox{$\sim$}~}\xspace}
\def\lsim{{~\raise.15em\hbox{$<$}\kern-.85em
          \lower.35em\hbox{$\sim$}~}\xspace}

\def\sqs   {\ensuremath{\protect\sqrt{s}}\xspace}

\def\ptot       {\mbox{$p$}\xspace}
\def\pt         {\mbox{$p_{\mathrm{ T}}$}\xspace}

\def\evtgen     {\mbox{\textsc{EvtGen}}\xspace}
\def\fewz       {\mbox{\textsc{Fewz}}\xspace}

\def\geant      {\mbox{\textsc{Geant4}}\xspace}

\def\photos     {\mbox{\textsc{Photos}}\xspace}
\def\powheg     {\mbox{\textsc{Powheg}}\xspace}
\def\pythia     {\mbox{\textsc{Pythia}}\xspace}

\def\tell1  {TELL1\xspace}
\def\ukl1   {UKL1\xspace}

\newcommand{\wj}{\ensuremath{\W{j}}\xspace}
\newcommand{\wpj}{\ensuremath{\Wp{j}}\xspace}
\newcommand{\wmj}{\ensuremath{\Wm{j}}\xspace}

\newcommand{\zj}{\ensuremath{\Z{j}}\xspace}

\newcommand{\muj}{\ensuremath{\mu\textrm{-jet}}\xspace}

\newcommand{\akt}{anti-$k_\text{T}$\xspace}

\newcommand{\cswpj}{\ensuremath{\sigma_{\wpj}}\xspace}
\newcommand{\cswmj}{\ensuremath{\sigma_{\wmj}}\xspace}
\newcommand{\cszj}{\ensuremath{\sigma_{\zj}}\xspace}
\newcommand{\ratiow}{\ensuremath{R_{\Wpm}}\xspace}
\newcommand{\ratiowz}{\ensuremath{R_{\W\Z}}\xspace}
\newcommand{\ratiowpz}{\ensuremath{R_{\Wp\Z}}\xspace}
\newcommand{\ratiowmz}{\ensuremath{R_{\Wm\Z}}\xspace}

\newcommand{\ptmu}{\ensuremath{p_{\textrm{T}}^{\mu}}\xspace}
\newcommand{\etamu}{\ensuremath{\eta^{\mu}}\xspace}
\newcommand{\ptmuj}{\ensuremath{p_{\textrm{T}}^{\muj}}\xspace}

\newcommand{\ptj}{\ensuremath{p_{\textrm{T}}^{\textrm{jet}}}\xspace}
\newcommand{\etaj}{\ensuremath{\eta^{\textrm{jet}}}\xspace}

\newcommand{\ptmjj}{\ensuremath{p_{\textrm{T}}^{\muj+j}}\xspace}
\newcommand{\ptmj}{\ensuremath{p_{\textrm{T}}^{\mu+j}}\xspace}

\newcommand{\mmm}{\ensuremath{M_{\mu\mu}}\xspace}

\newcommand{\rapz}{\ensuremath{y^{\Z}}\xspace}
\newcommand{\dphi}{\ensuremath{|\Delta\phi|}\xspace}

\def\mcatnlo     {\mbox{a\textsc{MC@NLO}}\xspace}

\usepackage{cite} % Allows for ranges in citations
\usepackage{mciteplus}

\usepackage{longtable} % only for template; not usually to be used in PAPERs

\usepackage[utf8]{inputenc}

\begin{document}

%%%%%%%%%%%%%%%%%%%%%%%%%
%%%%% Title     %%%%%%%%%
%%%%%%%%%%%%%%%%%%%%%%%%%
\renewcommand{\thefootnote}{\fnsymbol{footnote}}
\setcounter{footnote}{1}

% %%%%%%% CHOOSE TITLE PAGE--------
%\onecolumn
%\input{title-LHCb-INT}
%\input{title-LHCb-ANA}
%\input{title-LHCb-CONF}
% $Id: title-LHCb-PAPER.tex 91473 2016-05-03 12:17:38Z sfarry $
% ===============================================================================
% Purpose: LHCb-PAPER journal paper title page template
% Author: 
% Created on: 2010-09-25
% ===============================================================================

%%%%%%%%%%%%%%%%%%%%%%%%%
%%%%%  TITLE PAGE  %%%%%%
%%%%%%%%%%%%%%%%%%%%%%%%%
\begin{titlepage}
\pagenumbering{roman}

% Header ---------------------------------------------------
\vspace*{-1.5cm}
\centerline{\large EUROPEAN ORGANIZATION FOR NUCLEAR RESEARCH (CERN)}
\vspace*{1.5cm}
\noindent
\begin{tabular*}{\linewidth}{lc@{\extracolsep{\fill}}r@{\extracolsep{0pt}}}
\ifthenelse{\boolean{pdflatex}}% Logo format choice
{\vspace*{-2.7cm}\mbox{\!\!\!\includegraphics[width=.14\textwidth]{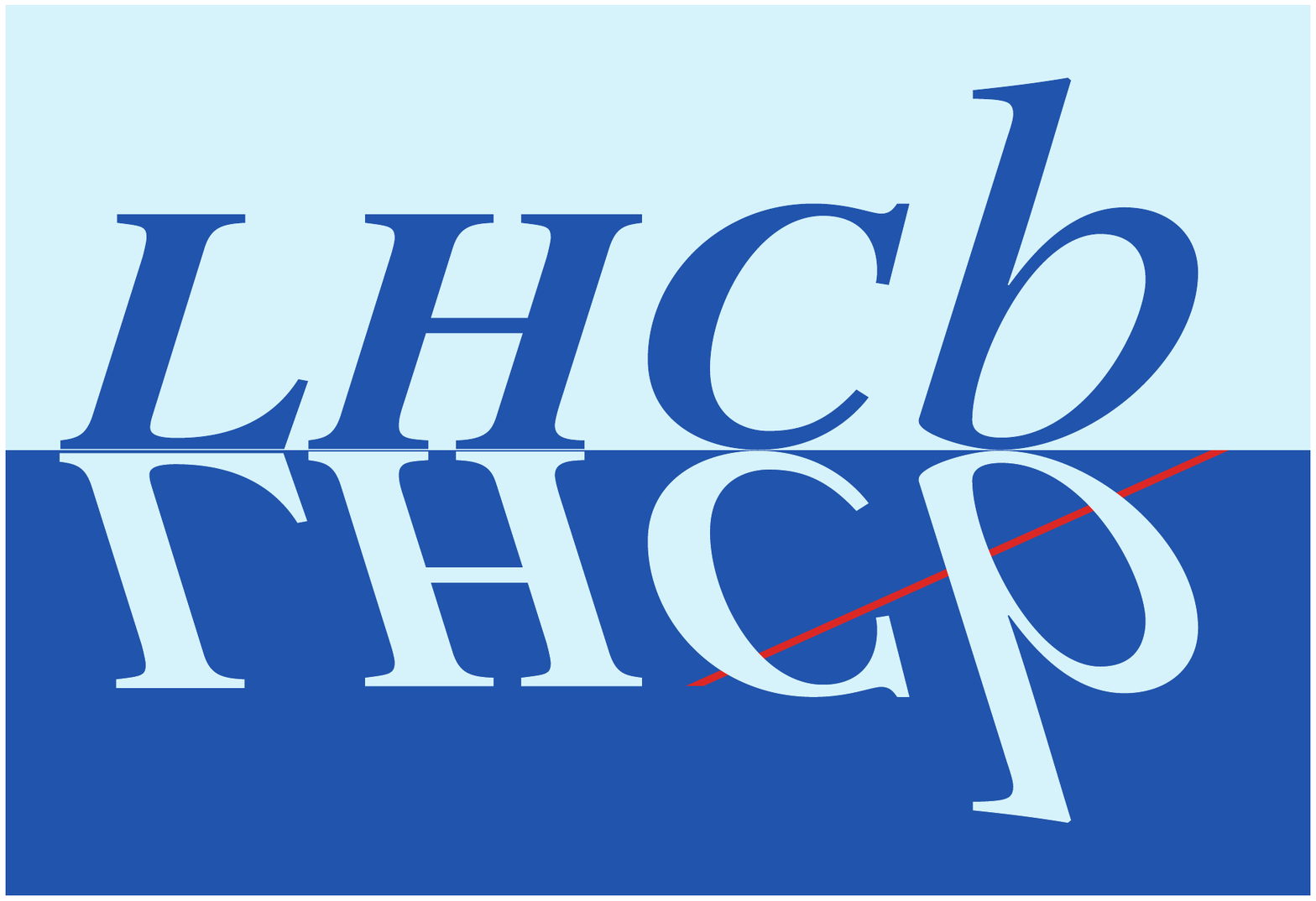}} & &}%
{\vspace*{-1.2cm}\mbox{\!\!\!\includegraphics[width=.12\textwidth]{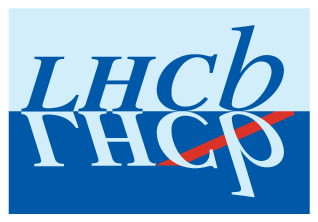}} & &}%
\\
 & & CERN-EP-2016-092 \\  % ID 
 & & LHCb-PAPER-2016-011 \\  % ID 
 & & 30 May 2016 \\ % Date - Can also hardwire e.g.: 23 March 2010
 & & \\
% not in paper \hline
\end{tabular*}

\vspace*{2cm}

% Title --------------------------------------------------
{\normalfont\bfseries\boldmath\huge
\begin{center}
  Measurement of forward \W \\ and \Z boson production \\in association with jets in proton-proton collisions at $\sqrt{s}=8$~TeV
\end{center}
}

\vspace*{1.3cm}

% Authors -------------------------------------------------
\begin{center}
%In the footnote, replace 'paper' by 'letter' in case of submission to PRL or PLB 
The LHCb collaboration\footnote{Authors are listed at the end of this paper.}
\end{center}

\vspace{\fill}

% Abstract -----------------------------------------------
\begin{abstract}
  \noindent The production of \W and \Z bosons in association with jets is studied in the forward region of proton-proton collisions collected at a centre-of-mass energy of 8~TeV by the LHCb experiment, corresponding to an integrated luminosity of \mbox{1.98 $\pm$ 0.02~fb$^{-1}$}.
%The \Z boson is reconstructed through its decay to muons, while the \W boson is reconstructed using its decay to a muon and a neutrinio.
The \W boson is identified using its decay to a muon and a neutrino, while the \Z boson is identified through its decay to a muon pair.
Total cross-sections are measured and combined into charge ratios, asymmetries, and ratios of $W+$jet and $Z$+jet production cross-sections.
Differential measurements are also performed as a function of both boson and jet kinematic variables.
All results are in agreement with Standard Model predictions.
  
\end{abstract}

\vspace*{1.0cm}

\begin{center}
Published in JHEP 05 (2016) 131
\end{center}

\vspace{\fill}

{\footnotesize 
\centerline{\copyright~CERN on behalf of the \lhcb collaboration, licence \href{http://creativecommons.org/licenses/by/4.0/}{CC-BY-4.0}.}}
\vspace*{2mm}

\end{titlepage}

%%%%%%%%%%%%%%%%%%%%%%%%%%%%%%%%
%%%%%  EOD OF TITLE PAGE  %%%%%%
%%%%%%%%%%%%%%%%%%%%%%%%%%%%%%%%

%  empty page follows the title page ----
\newpage
\setcounter{page}{2}
\mbox{~}
%\newpage
%
%% Author List ----------------------------
%%  You need to get a new author list!
%\input{LHCb_authorlist.tex}
%
%The author list for journal publications is provided by the Membership Committee shortly after 'approval to go to paper' has been given.
%%It will be made available on the page
%%\verb!http://www.physik.uzh.ch/~strauman/forMemCo/LHCb-PAPER-XXXX-XXX/! .
%It will be sent to you by email shortly after a paper number has beens assigned.
%The author list should be included already at first circulation, 
%to allow new members of the collaboration to verify whether they have been included correctly.
%Occasionally a misspelled name is corrected or associated institutions become full members.
%In that case, a new author list will be sent to you.
%In case line numbering doesn't work well after including the authorlist, try moving the \verb!\bigskip! after the last author to a separate line.
%
%
%The authorship for Conference Reports should be ``The LHCb
%  collaboration'', with a footnote giving the name(s) of the contact
%  author(s), but without the full list of collaboration names.

\cleardoublepage

%\twocolumn
% %%%%%%%%%%%%% ---------

\renewcommand{\thefootnote}{\arabic{footnote}}
\setcounter{footnote}{0}

%%%%%%%%%%%%%%%%%%%%%%%%%%%%%%%%
%%%%%  Table of Content   %%%%%%
%%%%%%%%%%%%%%%%%%%%%%%%%%%%%%%%
%%%% Uncomment next 2 lines if desired
%\tableofcontents
%\cleardoublepage

%%%%%%%%%%%%%%%%%%%%%%%%%
%%%%% Main text %%%%%%%%%
%%%%%%%%%%%%%%%%%%%%%%%%%

\pagestyle{plain} % restore page numbers for the main text
\setcounter{page}{1}
\pagenumbering{arabic}

%% Uncomment during review phase. 
%% Comment before a final submission.
%\linenumbers

% You can include short sections directly in the main tex file.
% However, for larger papers it is desirable to split the text into
% several semiautonomous files, which can be revised independently.
% This is especially useful when developing a document in
% collaboration with several people, since then different parts can be
% edited independently.  This type of file organization is shown here.
% 

%%%%%%%%%%%%%%%%%%%%%%%%%%%%%%%%%%%%
% !TEX root = main.tex
%%%%%%%%%%%%%%%%%%%%%%%%%%%%%%%%%%%%

\section{Introduction}
\label{sec:introduction}
Measurements of vector boson production in association with jets in the forward region at the Large Hadron Collider (LHC) can be used to test the Standard Model (SM) and provide constraints on the parton density functions (PDFs). LHCb is the only detector at the LHC with precision tracking coverage in the forward region, allowing sensitivity to PDFs at a different range of Bjorken-$x$ compared to ATLAS and CMS~\cite{Thorne:2008am}. LHCb measurements typically probe PDFs at $x$ as low as $10^{-4}$ and at high $x$~\cite{Farry:2015xha}.

This article reports total and differential cross-section measurements of \W and \Z production in association with jets, hereafter referred to as \wj and \zj, respectively.\footnote{Here, the notation \Z additionally includes contributions from virtual photon production and its interference with \Z boson production, $\Z/\gamma^*$.}  The measurements are performed using data collected during 2012 at a centre-of-mass energy of \sqs=8\tev, corresponding to an integrated luminosity of 1.98 $\pm$ 0.02~fb$^{-1}$. The \W and \Z bosons are identified through the $\W\to\mu\nu_\mu$ and $\Z\to\mu\mu$ decay channels. This work extends measurements of the \zj production cross-section at 7\tev~\cite{LHCb-PAPER-2013-058, LHCb-PAPER-2014-055} and ratios of the production cross-sections at 7 and 8\tev~\cite{LHCb-PAPER-2015-021}. It also complements previous studies of inclusive electroweak boson production at \lhcb, where the electroweak bosons decay to muons~\cite{LHCb-PAPER-2014-033,LHCb-PAPER-2015-001,LHCb-PAPER-2015-049}. 

This analysis makes use of the same fiducial acceptances for electroweak bosons as previously employed  in Ref.~\cite{LHCb-PAPER-2015-001}. For \W boson decays, this corresponds to requiring that the muon has a pseudorapidity, \etamu, in the range $2.0<\etamu<4.5$ and transverse momentum, \ptmu, greater than 20\gev.\footnote{This article uses natural units, where the speed of light ($c$) and the reduced Planck constant ($\hbar$) are set to unity, $c = \hbar = 1$.} For \Z boson decays, both muons are required to fulfil these kinematic requirements, and in addition, the dimuon invariant mass, \mmm, is required to be in the range $60<\mmm<120\gev$. The fiducial criteria for these measurements require at least one jet to have transverse momentum $\ptj > 20\gev$, and jet pseudorapidity, \etaj, in the range $2.2<\etaj<4.2$. The jet is also required to be separated by a radius $\Delta R$ of 0.5 from the charged lepton(s) produced in the boson decay, where $\Delta R$ is the sum in quadrature of the difference in pseudorapidity and the difference in azimuthal angle between the jet and the lepton. In addition, the \wj measurement requires that the transverse component of the vector sum of the muon and jet momenta, \ptmj, is greater than 20\gev. Jets are reconstructed using the \akt algorithm~\cite{Cacciari:2008gp}, with the $R$ parameter set to 0.5. Jet energies are defined at the hadron level, and do not include the contribution of neutrinos in the jet.

All measurements are performed for the jet with the largest transverse momentum in the event. The \wj measurement is made differentially as a function of \ptj, \etaj, and the pseudorapidity of the muon produced by the \W boson decay, \etamu.  For the \zj measurement, the differential cross-sections are determined as a function of \ptj, \etaj, the boson rapidity, \rapz, and the difference in azimuthal angle between the \Z boson and the jet, \dphi. The jet transverse momentum distributions and the \dphi distribution tend to be sensitive to higher-order effects within perturbative quantum chromodynamics (QCD)~\cite{Alioli:2010xd}, while measurements of the (pseudo)rapidity distributions are sensitive to the PDFs that parameterise the structure of the proton. The ratio of the \wpj to the \wmj cross-sections is measured, as is the ratio of the \wj cross-sections to the \zj cross-section. Finally, the charge asymmetry of \wj production is measured as a function of \etamu.

%%%%%%%%%%%%%%%%%%%%%%%%%%%%%%%%%%%%
% !TEX root = main.tex
%%%%%%%%%%%%%%%%%%%%%%%%%%%%%%%%%%%%
\section{Detector and simulation}
\label{sec:Detector}
The \lhcb detector~\cite{Alves:2008zz,LHCb-DP-2014-002} is a single-arm forward
spectrometer covering the \mbox{pseudorapidity} range $2<\eta <5$,
designed for the study of particles containing \bquark or \cquark
quarks. The detector includes a high-precision tracking system
consisting of a silicon-strip vertex detector surrounding the $pp$
interaction region, a large-area silicon-strip detector located
upstream of a dipole magnet with a bending power of about
$4{\mathrm{\,Tm}}$, and three stations of silicon-strip detectors and straw
drift tubes placed downstream of the magnet.
The tracking system provides a measurement of momentum, \ptot, of charged particles with
a relative uncertainty that varies from 0.5\% at low momentum to 1.0\% at 200\gev.
The minimum distance of a track to a primary vertex (PV), the impact parameter, is measured with a resolution of $(15+29/\pt)\mum$,
where \pt is the component of the momentum transverse to the beam, in\,\gev.
Different types of charged hadrons are distinguished using information
from two ring-imaging Cherenkov detectors. 
Photons, electrons and hadrons are identified by a calorimeter system consisting of
scintillating-pad (SPD) and preshower detectors, an electromagnetic
calorimeter and a hadronic calorimeter. Muons are identified by a
system composed of alternating layers of iron and multiwire
proportional chambers.
The online event selection is performed by a trigger, 
which consists of a hardware stage, based on information from the calorimeter and muon
systems, followed by a software stage, which applies a full event
reconstruction.

In this paper, candidate events are required to pass the hardware trigger,
  which selects muons with a transverse momentum $\pt>1.76\gev$ and the subsequent software trigger, where a muon with 
  $\pt>10\gev$ is required to be present. A global event cut (GEC) is also applied at the hardware stage, which requires that the number of hits in the SPD sub-detector should be less than 600.

Simulated $pp$ collisions are generated using
\pythia8~\cite{Sjostrand:2007gs,*Sjostrand:2006za} 
 with a specific \lhcb
configuration~\cite{LHCb-PROC-2010-056}.  Decays of hadronic particles
are described by \evtgen~\cite{Lange:2001uf}, in which final-state
radiation is generated using \photos~\cite{Golonka:2005pn}. The
interaction of the generated particles with the detector, and its response,
are implemented using the \geant
toolkit~\cite{Allison:2006ve, *Agostinelli:2002hh} as described in
Ref.~\cite{LHCb-PROC-2011-006}.

Results are compared to theoretical calculations performed at $\mathcal{O}(\alpha_s^2)$ in perturbative QCD using the \powheg~\cite{Alioli:2010xd, Alioli:2010qp} and \mcatnlo~\cite{Alwall:2014hca} generators, interfaced with \pythia in order to simulate the parton shower, where the NNPDF3.0~\cite{Ball:2010de,Ball:2014uwa} PDF set is used to describe the dynamics of the colliding protons. Additional fixed-order predictions are generated using \fewz~\cite{Gavin:2010az} at $\mathcal{O}(\alpha_s^2)$ with the NNPDF3.0, CT14~\cite{Dulat:2015mca} and MMHT14~\cite{Harland-Lang:2014zoa} PDF sets.

%%%%%%%%%%%%%%%%%%%%%%%%%%%%%%%%%%%%
% !TEX root = main.tex
%%%%%%%%%%%%%%%%%%%%%%%%%%%%%%%%%%%%

\section{Event selection}
\label{sec:selection}
Events are selected containing one or two high-\pt muons produced in association with a high-\pt jet. Jets are reconstructed at LHCb using a particle flow algorithm~\cite{LHCb-PAPER-2013-058} and clustered using the \akt algorithm as implemented in {\sc Fastjet}~\cite{Cacciari:2005hq}.  Additional selection requirements are placed on the jet properties in order to reduce the number of spurious jets selected. The jet energies are calibrated on an event-by-event basis. These calibrations are determined from both data and simulation, and are applied as a function of the jet \pt, azimuthal angle, pseudorapidity, charged particle fraction and the number of reconstructed PVs in the event~\cite{LHCb-PAPER-2013-058}. To reduce contamination from multiple $pp$ interactions, charged particles reconstructed within the vertex detector are only clustered into a jet if they are associated to the same PV as the final state muon(s).

The measured muons and jets are required to satisfy the fiducial requirements outlined in Sec.~\ref{sec:introduction}. An exception is the requirement on the \pt of the vector sum of the momentum of the muon and jet,  $\ptmj>20\gev$, in \wj events. In  the selection, the muon is replaced by the jet, \muj, which contains the signal muon after performing a jet reconstruction with relaxed jet selection requirements. The modified fiducial requirement, $\ptmjj>20\gev$, improves the suppression of the background from di-jets, which tend to be balanced in transverse momentum. An acceptance factor is introduced (see Sec.~\ref{sec:xsec}), which corrects the results to correspond to the fiducial regions defined in Sec.~\ref{sec:introduction}.

As \wj events contain just one final-state muon and consequently suffer from a higher background, additional requirements are placed on the sample. The background to the \wj sample from \zj events where both muons are produced in the LHCb acceptance is suppressed by rejecting events containing a second muon with \pt in excess of 20\gev. Backgrounds from semileptonic decays of heavy-flavour hadrons are suppressed by requiring that the impact parameter of the muon track with respect to the PV should be less than 0.04\mm.  Additionally, the sum of the energy associated with the track in the electromagnetic and hadronic calorimeters is required to be less than 4\% of the muon momentum. In total, 8\,162 \zj and 133\,746 (99\,683) \wpj (\wmj) candidates are selected.

%%%%%%%%%%%%%%%%%%%%%%%%%%%%%%%%%%%%
% !TEX root = main.tex
%%%%%%%%%%%%%%%%%%%%%%%%%%%%%%%%%%%%
\section{Purity determination}
The selected data samples contain background contributions from three distinct processes:
\begin{itemize}
\item QCD multi-jet production, which can produce muons in the final state, either due to the misidentification of hadrons, or through the semileptonic decay of heavy-flavour hadrons where a high-\pt jet is also present in the event.
\item Electroweak processes, such as $Z\to\tau\tau$, $W\to\tau\nu$ or, in the case of \wj production, $Z\to\mu\mu$, can produce events that mimic the signal. Contributions are also expected from electroweak diboson and top quark production.
\item A small background contribution from ``fake jets'' is present when the data sample contains events where the reconstructed and identified jet is not associated with genuine particles, but is instead due to detector effects, such as the presence of fake or misreconstructed particles, or to particles produced in a different $pp$ collision to that producing the $W$ or $Z$ boson.

\end{itemize}

\subsection{\wj sample purity}
\label{sec:wpurity}
The QCD background to the \wj sample is determined by performing an extended maximum likelihood fit to the distribution of the muon transverse momentum \ptmu, divided by the transverse momentum of the \muj, \ptmuj (where the \muj is defined in Sec.~\ref{sec:selection}). This variable acts as a measure of muon isolation, with a value close to unity when little activity is present in the vicinity of the candidate muon and a value closer to zero as the multiplicity in the surrounding region increases. Consequently, it provides strong discrimination between muons produced in electroweak processes, which tend to be produced in isolation, and those produced in QCD processes, which are typically surrounded by additional particles. Two separate components are accounted for in the fit:
\begin{itemize}
\item The template shape describing all electroweak processes, including the signal, is taken from simulation. The shape of the isolation variable is approximately independent of \ptmu, and consequently provides a good description of all electroweak processes. The simulated shape is corrected for mismodelling by applying correction factors obtained from a comparison of \zj events in data and simulation. The \wj signal contribution is subsequently separated from the other electroweak backgrounds as described below.
\item The QCD background template is obtained using a di-jet enriched data sample, obtained by requiring $\ptmjj < 20\gev$. The small contribution from signal events in the template is subtracted using simulation where the normalisation is obtained from the bin corresponding to $\ptmu/\ptmuj>0.95$ in the signal region. The template shape is then corrected for differences in the \ptmuj distribution between the background and signal regions.
\end{itemize}
The fits are performed in bins of \etaj, \ptj, and \etamu separately for positively and negatively charged \wj candidates. The background from \Z decays to muons and $\tau$ leptons, where a single muon is present in the final state, is determined from simulation where the sample is normalised to the number of fully reconstructed $\Z\to\mu\mu$ decays observed in data.  The small contribution from $WW$, $t\bar{t}$ and single top events is determined using next-to-leading order (NLO) predictions obtained from MCFM~\cite{Campbell:2000bg}. Finally, the background from $\W\to\tau\nu$ decays is determined by first obtaining the ratio of $W\to\tau\nu$ to $W\to\mu\nu$ events expected from simulation and normalising to the remaining signal after all other backgrounds have been determined. The background from fake jets is evaluated using simulation.

\begin{figure}[t!]
\begin{center}
\includegraphics[width=0.49\textwidth]{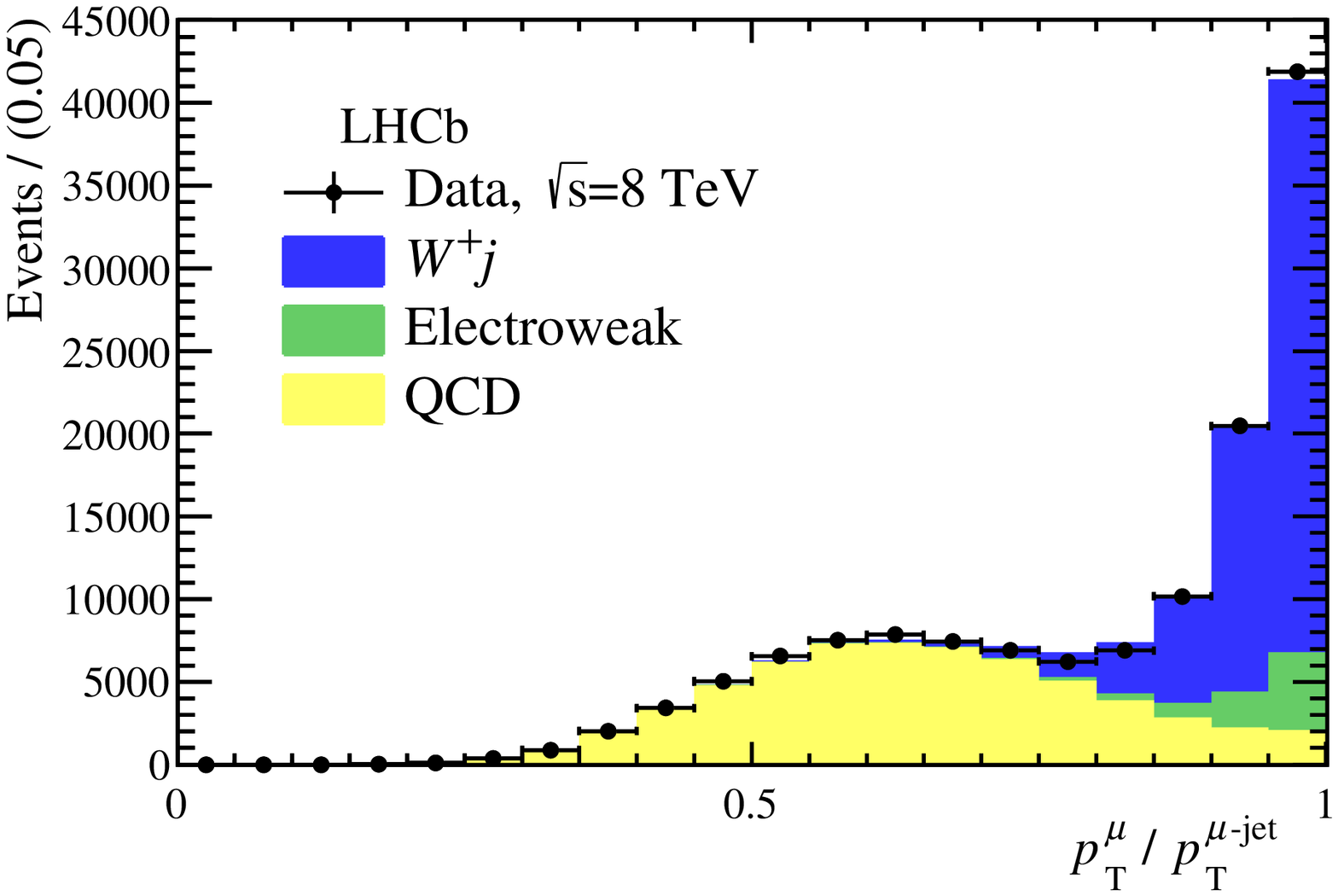}
\includegraphics[width=0.49\textwidth]{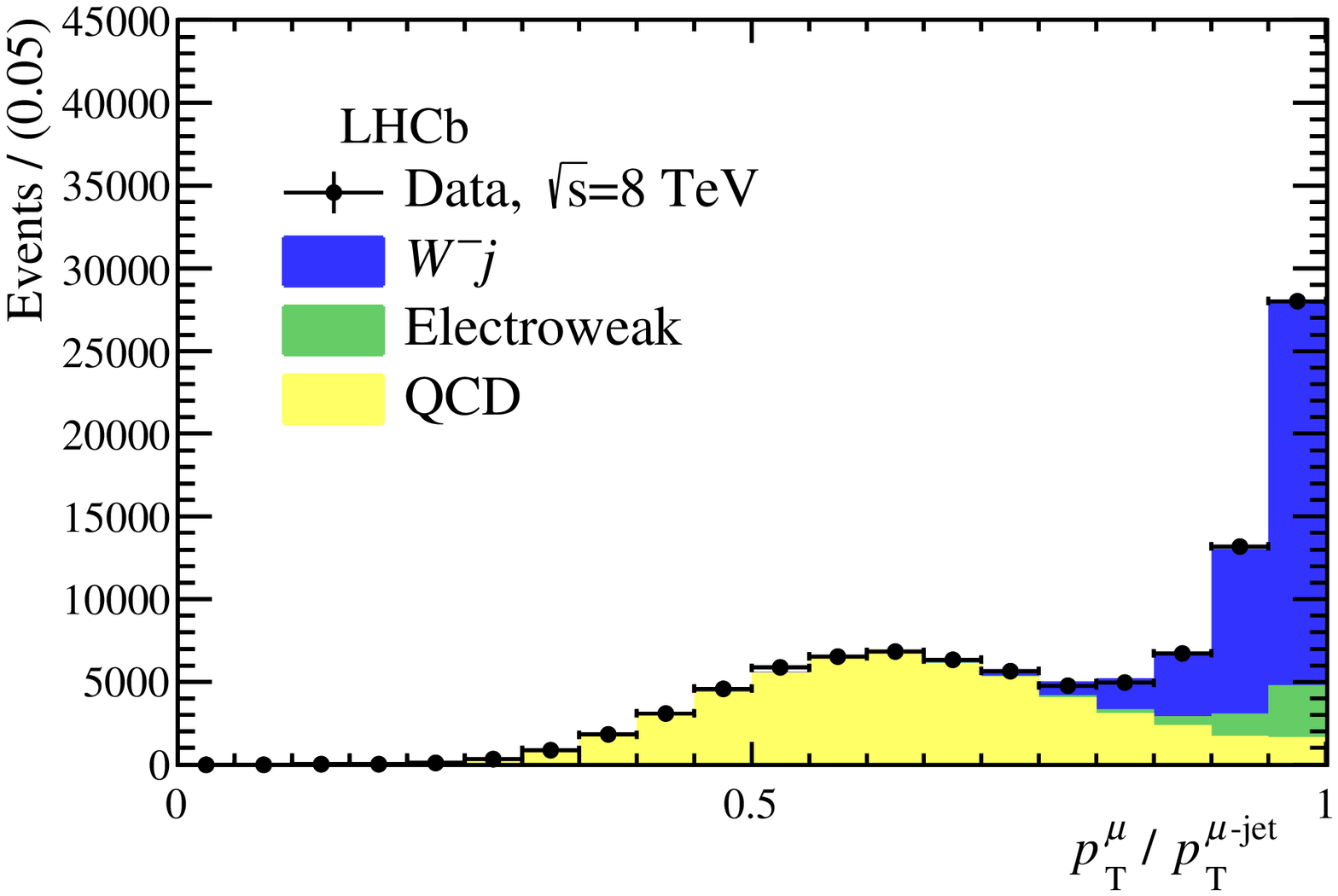}
\caption{The contributions to the selected (left) \wpj and (right) \wmj samples are shown, where the QCD background is obtained by a fit to the \ptmu/\ptmuj spectrum and the electroweak background is determined as described in the text. The contributions shown are the sum of the individual contributions in bins of \etaj, where the charge asymmetry typical of \wj production in $pp$ collisions is evident.}
\label{fig:wpurity}
\end{center}
\end{figure}

The contribution from QCD processes is found to vary between 30--70\% in different bins of \etaj, \ptj and \etamu while the contribution from electroweak processes (including di-boson and top production) amounts to 5--10\% of the selected samples. The contribution from fake jets represents approximately 0.8--0.9\% of the samples. The overall purity of the \wpj(\wmj) sample is determined to be 46.7(36.5)\% where the total contributions, obtained by summing over the yields in the \etaj bins, are shown in Fig.~\ref{fig:wpurity}.

\subsection{\zj sample purity}
\label{sec:zjpurity}
The contribution from semileptonic decays of heavy-flavour particles to the \zj sample is determined by selecting a background-enhanced sample using two approaches, where either the muons are not isolated from other activity in the event or where they do not form a good vertex. The efficiency with which the requirements select background events is evaluated by comparing the number of events selected by the two approaches as in Ref.~\cite{LHCb-PAPER-2015-001}. The total contribution is estimated to be approximately 0.7\%. The misidentification of hadrons as muons is evaluated as in Ref.~\cite{LHCb-PAPER-2015-001}, by considering the contribution from events where both muons fulfil all the selection criteria, but with both muons required to have the same sign charge; and gives a contribution of approximately 0.4\%. Decays of the \Z boson to $\tau$ pairs can contribute if both $\tau$ leptons subsequently decay to muons. The contribution from this source is determined from simulation to be approximately 0.1\%. The number of events containing di-boson or top production is again calculated using simulation, normalised to NLO predictions from MCFM and is determined to be negligible. The contribution from fake jets is determined from simulation to amount to approximately 0.9\% of the selected sample.  The overall purity of the \zj sample is determined to be 97.8\%.

%%%%%%%%%%%%%%%%%%%%%%%%%%%%%%%%%%%%
% !TEX root = main.tex
%%%%%%%%%%%%%%%%%%%%%%%%%%%%%%%%%%%%
\section{Cross-section measurement}
\label{sec:xsec}
The cross-section, $\sigma_i$, for \W and \Z boson production in association with one or more jets in the $i^{th}$ phase space bin is given by
\begin{equation}
\sigma_{i} = U_i\frac{\mathcal{A}_i\cdot\rho_i\cdot N_i}{\varepsilon^{\rm muon}_i\cdot\varepsilon^{\rm jet}_i\cdot\varepsilon^{\rm sel}_i\cdot\mathcal{L}} \, ,
\end{equation}
where $U_i$ is an unfolding correction which accounts for resolution effects causing migrations between different bins of phase space. The number of candidates selected in bin $i$ is given by $N_i$ while $\rho_i$ represents the signal purity. The acceptance factor, $\mathcal{A}_{i}$, accounts for differences between the fiducial region of the measurement and the kinematic requirements placed on the muons and jets. The efficiencies for reconstructing the muons and the jet are given by $\varepsilon^{\rm muon}_i$ and $\varepsilon^{\rm jet}_i$, respectively, while the efficiency of any additional event selection is given by $\varepsilon^{\rm sel}_i$.

The instantaneous luminosity is measured continuously during the acquisition of physics data by recording the rates of several selected standard processes.
The effective absolute cross-section of these processes is measured during dedicated calibration periods, using both van der Meer scans~\cite{vanderMeer:1968zz, Barschel:1693671} and beam-gas imaging methods specific to the LHCb detector~\cite{FerroLuzzi:2005em}.
Both methods give consistent results and are combined to give the final luminosity calibration with an uncertainty of 1.2\%~\cite{LHCb-PAPER-2014-047}.
The integrated luminosity of the data sample used, $\mathcal{L}$, is obtained from the accumulated counts of the calibrated rates and amounts to 1.98 $\pm$ 0.02~fb$^{-1}$.

%The absolute luminosity is measured during dedicated data taking periods, using both Van der Meer scans\cite{vanderMeer:1968zz} and beam-gas imaging methods~\cite{FerroLuzzi:2005em, Barschel:1693671}.
%An effective cross-section of selected standard processes is calibrated during those measurements. During normal data taking periods the interaction rate of the calibrated processes is continuously measured.
%The integrated luminosity of any data sample, $\mathcal{L}$ is then obtained from the accumulated counts of a calibrated visible cross-section. Both methods give consistent results and are combined to give the final luminosity estimate with an uncertainty of 1.16\%~\cite{LHCb-PAPER-2014-047}. The luminosity of the sample used is determined to be 1.98$\pm$0.02~fb$^{-1}$.

The efficiency to reconstruct and select muons in the event is evaluated using the same techniques employed in the inclusive \W and \Z boson measurements at LHCb~\cite{LHCb-PAPER-2014-033,LHCb-PAPER-2015-001,LHCb-PAPER-2015-049}. In particular, a data-driven tag-and-probe study is performed on selected inclusive $\Z\to\mu\mu$ events in data and the efficiency of reconstructing, triggering and identifying the muons is measured. These efficiencies are applied as a function of the pseudorapidity of the muon(s) in the event. The efficiency to reconstruct and identify the jet in the event $\varepsilon^{\rm jet}_i$, is evaluated from simulation. This efficiency increases with \pt, from about 90\% for jets with \pt of 20\gev to saturate at about 95\% for higher \pt jets. It is dominated by the probability that the jet passes the requirements designed to reject fake jets. In the case of the \wj sample, the efficiency of the additional requirements placed on the event, including a veto on extra muons, is evaluated using a ``pseudo-\wj'' sample, where \zj events are selected but one muon is masked in order to mimic the neutrino in \wj events. Corrections are applied based on a comparison of the efficiency of the requirements in \wj and ``pseudo-\wj'' events in simulation. The efficiency of the GEC requirement at the hardware stage of the trigger is again evaluated in a similar fashion to the inclusive analyses, where the efficiency is measured in a \zj sample selected with a looser trigger requirement~\cite{LHCb-PAPER-2014-033,LHCb-PAPER-2015-001,LHCb-PAPER-2015-049}. This efficiency is evaluated separately in each kinematic bin considered in the analysis, but shows little variation with the variables that describe the jet kinematics.

The unfolding correction, $U_i$, corrects for differences observed in the number of events produced and measured in a given bin due to the finite resolution of the detector, where the differences are primarily caused by migrations in the \ptj and \etaj distributions. The correction is determined from simulation as the ratio of events produced in a specific bin to those recorded by the detector in the same bin. The correction varies between 0.9 and 1.0, where the largest corrections are seen at low \ptj and in the highest and lowest \etaj bins.

For the \zj sample, the acceptance factor, $\mathcal{A}_i$, is identically equal to unity as the selection mirrors the fiducial acceptance exactly. In the case of the \wj selection, the requirement of \mbox{$\ptmjj > 20\gev$} differs from the fiducial requirement of $\ptmj > 20\gev$. Consequently, the acceptance factor accounts for differences between these two variables arising from extra activity that may be present in the neighbourhood of the signal muon. This factor is evaluated using simulation, which is reweighted in bins of jet \pt and pseudorapidity to match next-to-leading order predictions obtained from \mcatnlo. The acceptance factor varies between 0.95 and 1.00 in different bins of phase space.

\section{Systematic uncertainties}
Several sources of systematic uncertainty have been evaluated. The uncertainty on the estimated purity of the \wj sample is evaluated by repeating the fit using alternative templates. The fit is performed for a number of different scenarios:
\begin{itemize}
\item the data-driven corrections are not applied to the simulated \wj shape,
\item the simulated \wj shape is replaced by the ``pseudo-\wj'' data sample,
\item the subtraction of signal events from the background template is performed by obtaining the normalisation from simulation instead of the data-driven method outlined in Sec.~\ref{sec:wpurity}.
\end{itemize}
The uncertainty on the contributions from electroweak templates is taken to be the statistical precision on the \zj and \wj samples used to perform the data-driven normalisation. For the \zj sample, the uncertainty on the misidentification background is given by the sum in quadrature of the statistical precision and the accuracy of the method, obtained by comparing the two approaches described in Sec.~\ref{sec:zjpurity}. This gives an uncertainty of approximately 30\% on the misidentification background. The uncertainty on the contribution from semileptonic decays of heavy-flavour hadrons is about 20\%, consisting of the sum in quadrature of the statistical uncertainty on the evaluated contribution, and the variation in the background level found by changing the requirements used in selecting the background-enhanced region. The uncertainty due to the presence of fake jets is taken to be the statistical uncertainty of approximately 30\% on the determination of the fake-jet contribution. A similar level of agreement is observed between data and simulation by comparing kinematic distributions in regions with enhanced fake-jet populations.

The uncertainty in the muon reconstruction efficiency is determined by re-evaluating the cross-section with the total efficiency varied by one standard deviation around the central value. An additional 1\% systematic uncertainty is also applied to account for differences in efficiencies observed between inclusive \Z events and \zj events. The uncertainty on the jet reconstruction efficiency is evaluated by comparing the differences in efficiency between \zj data and simulation where the quality requirements are varied about their nominal values. This results in an uncertainty of 1.9\%.
The uncertainty on the selection efficiency, 1\%, includes the statistical uncertainty due to the limited size of the ``pseudo-\wj'' data sample and the uncertainty on the corrections evaluated from simulation for differences between \wj and ``pseudo-\wj'' events.
The uncertainty on the GEC efficiency is taken to be the sum in quadrature of the accuracy of the method, 0.3\%~\cite{LHCb-PAPER-2015-001,LHCb-PAPER-2015-049}, and the difference observed between \wpj, \wmj and \zj events in simulation, typically smaller than 0.2\%. The uncertainty on the efficiency with which jets are selected is evaluated by varying the selection requirements and determining how the fraction of events rejected agrees between data and simulation, using the methods described in Ref.~\cite{LHCb-PAPER-2013-058}. Agreement is typically seen at the level of about 1.7\%. This is taken as an uncertainty on the modelling of the efficiencies in simulation, and is combined in quadrature with the statistical precision with which the efficiencies are determined.

The uncertainty on the acceptance factor, $\mathcal{A}_{i}$, is determined by comparing the values obtained with and without NLO reweighting performed, and by comparing the acceptance calculated in ``pseudo-\wj'' events in data and simulation. These individual differences, contributing 0.5\% and 0.3\%, respectively, are added in quadrature with the statistical precision of the determination.

Two contributions to the uncertainty on the unfolding correction, $U_{i}$, are considered. The variation of the corrections is evaluated by comparing the difference in the number of \zj events between the bin-by-bin corrections employed in the analysis and a Bayesian unfolding~\cite{DAgostini:1994zf,Adye:2011gm} with two iterations. The difference is typically 0.8--1.5\%, depending on the distribution considered. This is larger than the variation seen when changing the number of iterations in the Bayesian approach, and it is also larger than the effect of reweighting the bin-by-bin corrections to the jet transverse momentum distributions produced by different event generators. An additional uncertainty due to the resolution of the jet pseudorapidity in data is also considered and obtained by comparing the difference between the jet pseudorapidity calculated using just the charged component of the jet and using both the charged and neutral components in \zj data and simulation. A good level of agreement is observed within the statistical precision of 0.5\%. The two contributions are added in quadrature and taken as the systematic uncertainty associated with the unfolding corrections.

Different sources for the jet energy scale uncertainty are considered. The energy scale associated with tracks is known and simulated to an accuracy of better than 1\%~\cite{LHCb-DP-2014-002}. The calorimeter energy scales are modelled to an accuracy of better than 10\%. This is confirmed by comparing the fraction of \ptj carried by neutral final-state particles between data and simulation, and evaluating how much the calorimeter response can be varied before disagreement is observed. The jet energy resolution at LHCb is modelled in simulation to an accuracy of about 10\%~\cite{LHCb-PAPER-2013-058,LHCb-PAPER-2015-021}. The analysis is repeated with the simulated \ptj smeared by 10\%; the change in the final result of approximately 0.3\% is assigned as the relevant uncertainty. Combining these effects yields an energy scale uncertainty of about 3\%, consistent with previous studies~\cite{LHCb-PAPER-2013-058} considering the \pt balance in \Z+1-jet events. In order to determine the effect on the measurement, the analysis is repeated with the energy scale varied to cover possible differences between data and simulation. The variation in the measured cross-sections lies between 4 and 11\%, depending on the bin and sample considered. This is assigned as the energy scale uncertainty.

A summary of the different contributions to the systematic and total uncertainty for the measured quantities which will be outlined in Sec.~\ref{sec:results} is given in Table~\ref{tab:syst_uncerts}. In the case of \zj measurements, the systematic uncertainty is dominated by the knowledge of the jet energy scale, while for \wj measurements a similarly large uncertainty is present due to the determination of the sample purity.

%The jet energy scale was previously studied in Ref.~\cite{LHCb-PAPER-2013-058}, with the dominant contribution to the uncertainty being estimated from comparisons between data and simulation of the \pt balance between \PZ bosons and jets in \PZ+1-jet events. The typical uncertainty from these studies was about 3\% for jets with $\ptj > 20\gev$. The jet energy scale uncertainty is evaluated again here using a novel approach where different sources are considered separately. The energy scale associated with tracks is known and simulated to an accuracy of better than 1\%~\cite{LHCb-DP-2014-002}. The calorimeter energy scales are modelled to an accuracy of better than 10\%. This is confirmed by comparing the fraction of \ptj carried by neutral final-state particles between data and simulation, and evaluating how much the calorimeter response can be varied before disagreement is observed. Combining these effects yields an energy scale uncertainty consistent with previous studies. In order to determine the effect on the measurement, the analysis is repeated with the energy scale varied to cover possible differences between data and simulation. The variation in the measured cross-sections lies between 4 and 11\%, depending on the bin and sample considered. This is assigned as the energy scale uncertainty.

%%%%%%%%%%%%%%%%%%%%%%%%%%%%%%%%%%%%
% !TEX root = main.tex
%%%%%%%%%%%%%%%%%%%%%%%%%%%%%%%%%%%%

\section{Results}
\label{sec:results}
The total cross-sections for \wj and \zj production are obtained by summing over the measured cross-sections in bins of \etaj. All statistical uncertainties are taken to be uncorrelated, while uncertainties arising from common sources and/or methods are taken to be fully correlated between different bins. The cross-sections are calculated to be
\begin{eqnarray*}
\cswpj  &=& 56.9 \pm  0.2  \phantom{0}\pm 5.1  \phantom{0}\pm 0.7 \pb \, , \\
\cswmj &=& 33.1 \pm  0.2  \phantom{0}\pm 3.5  \phantom{0}\pm 0.4 \pb \, , \\
\cszj    &=& 5.71 \pm  0.06  \pm 0.27  \pm 0.07 \pb \, ,
\end{eqnarray*}
where the first uncertainties are statistical, the second are systematic, and the third are due to the luminosity determination. The ratios of \wj and \zj production are determined to be
\begin{eqnarray*}
\ratiowz &=& 15.8\pm  0.2  \phantom{0}\pm 1.1 \, , \\ 
\ratiowpz &=& 10.0 \pm  0.1  \phantom{0}\pm 0.6 \, , \\ 
\ratiowmz &=& 5.8 \phantom{0}\pm  0.1  \phantom{0}\pm 0.5 \, , \\ 
\ratiow &=& 1.72 \pm  0.01  \pm 0.06 \, ,
\end{eqnarray*}
where \ratiowz, \ratiowpz and \ratiowmz represent, respectively, the ratio of the \wj, \wpj and \wmj cross-sections to the \zj cross-section, and \ratiow represents the ratio of the \wpj to \wmj cross-sections. The asymmetry of \wpj and \wmj production, $A(\wj)$, is given by
\begin{eqnarray*}
A(\wj) \equiv (\cswpj - \cswmj)/(\cswpj + \cswmj ) &=& 0.264 \pm  0.003  \pm 0.015 \,.
\end{eqnarray*}
In the above results, the first uncertainties are statistical and the second are systematic.

\begin{table}[t!]
\begin{center}
\caption{Summary of the different contributions to the total uncertainty on \cswpj, \cswmj, \cszj and their ratios given as a percentage of the measured observable.}
\label{tab:syst_uncerts}
\begin{tabular}{c c c c  c c} 
Source & \cswpj & \cswmj & \cszj & \ratiowz & \ratiow \\ 
\hline 
Statistical & 0.4 & 0.5 & 1.1  & 1.2 & 0.7 \\ 
\hline 
Muon reconstruction & 1.3 & 1.3 & 0.6  & 0.9 & 0.0 \\ 
Jet reconstruction & 1.9 & 1.9 & 1.9 & 0.0 & 0.0  \\ 
Selection & 1.0 & 1.0 & 0.0 & 1.0 & 0.0  \\ 
GEC & 0.5 & 0.5 & 0.4 & 0.2 & 0.1  \\ 
Purity & 5.5 & 7.0 & 0.4 & 6.0 & 2.5  \\ 
Acceptance & 0.6 & 0.6 & 0.0 & 0.6 & 0.0  \\ 
Unfolding & 0.8 & 0.8 & 0.8 & 0.0 & 0.2  \\ 
Jet energy   & 6.5 & 7.7 & 4.3 & 3.4 & 1.2  \\ 
\hline 
 Total Systematic & 8.9 & 10.7 & 4.8 & 7.0 & 3.3  \\ 
\hline 
Luminosity & 1.2 & 1.2 & 1.2 & -- & --  \\ 
\end{tabular} 
\end{center}
\end{table}

\begin{figure}[ht!]
\begin{center}
\includegraphics[width=0.8\textwidth]{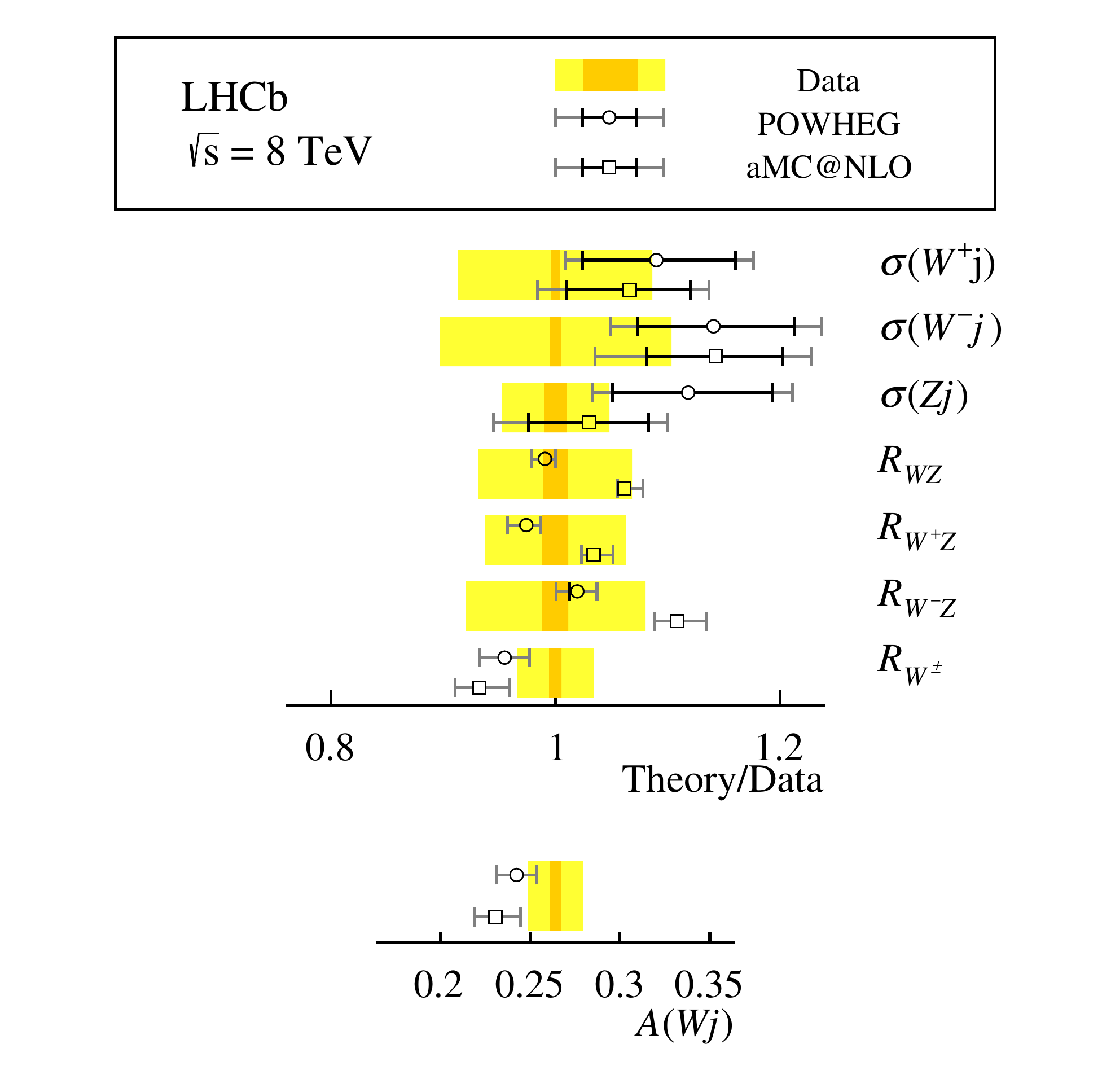}
\caption{\small Summary of the measurements performed in the fiducial region, as defined in Sec.~\ref{sec:introduction}. The measurements are shown as bands, while the theoretical predictions are presented as points. For the experimental measurements, the inner band represents the statistical uncertainty, while the outer band represents the total uncertainty. For the theory points, the inner error bar represents the scale uncertainty, while the outer bar represents the total uncertainty. The cross-sections and ratios are shown normalised to the measurement, while the asymmetry is presented separately.}
\label{fig:summary}
\end{center}
\end{figure}

\begin{figure}[h!]
\begin{center}
\includegraphics[width=0.495\textwidth]{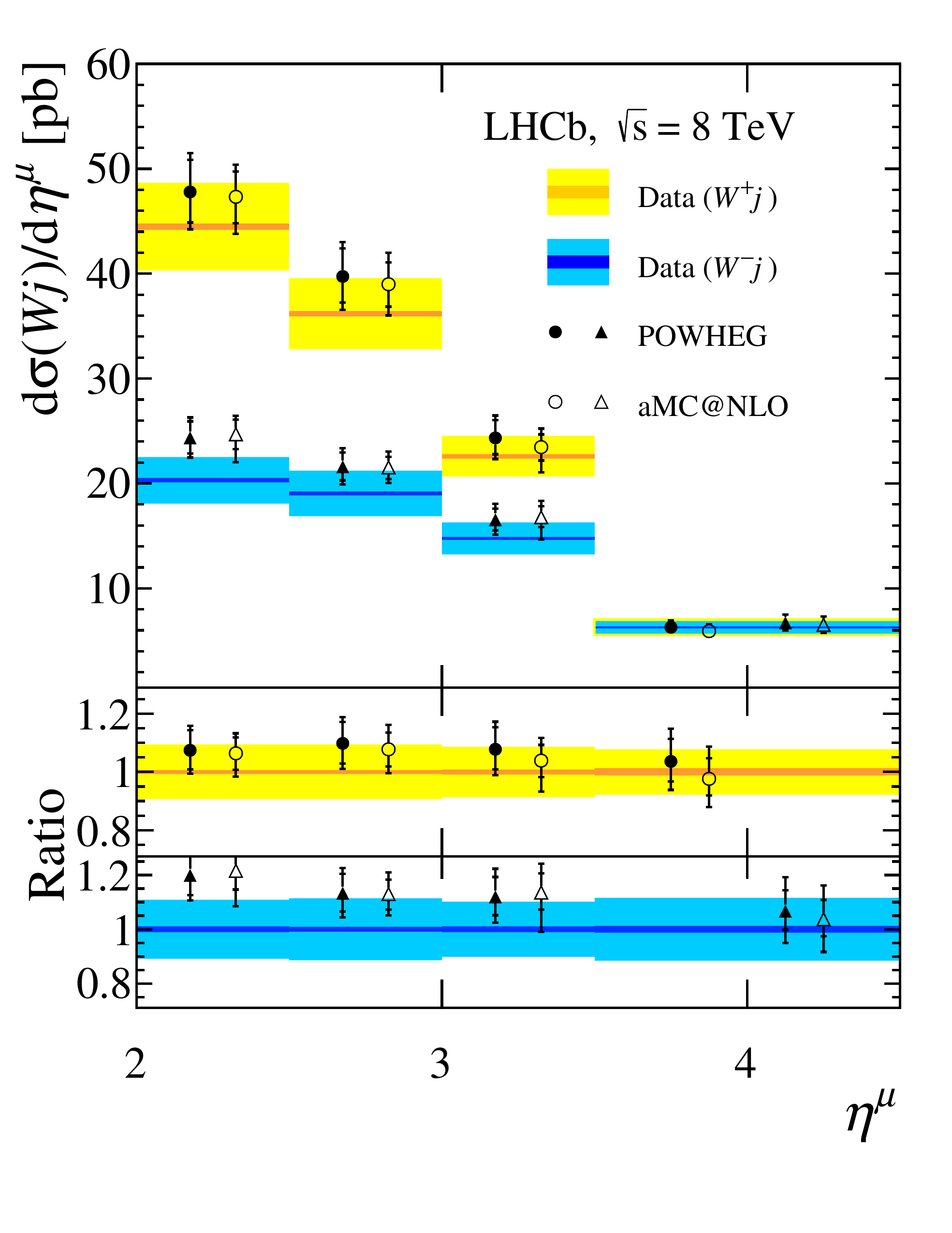}
\includegraphics[width=0.495\textwidth]{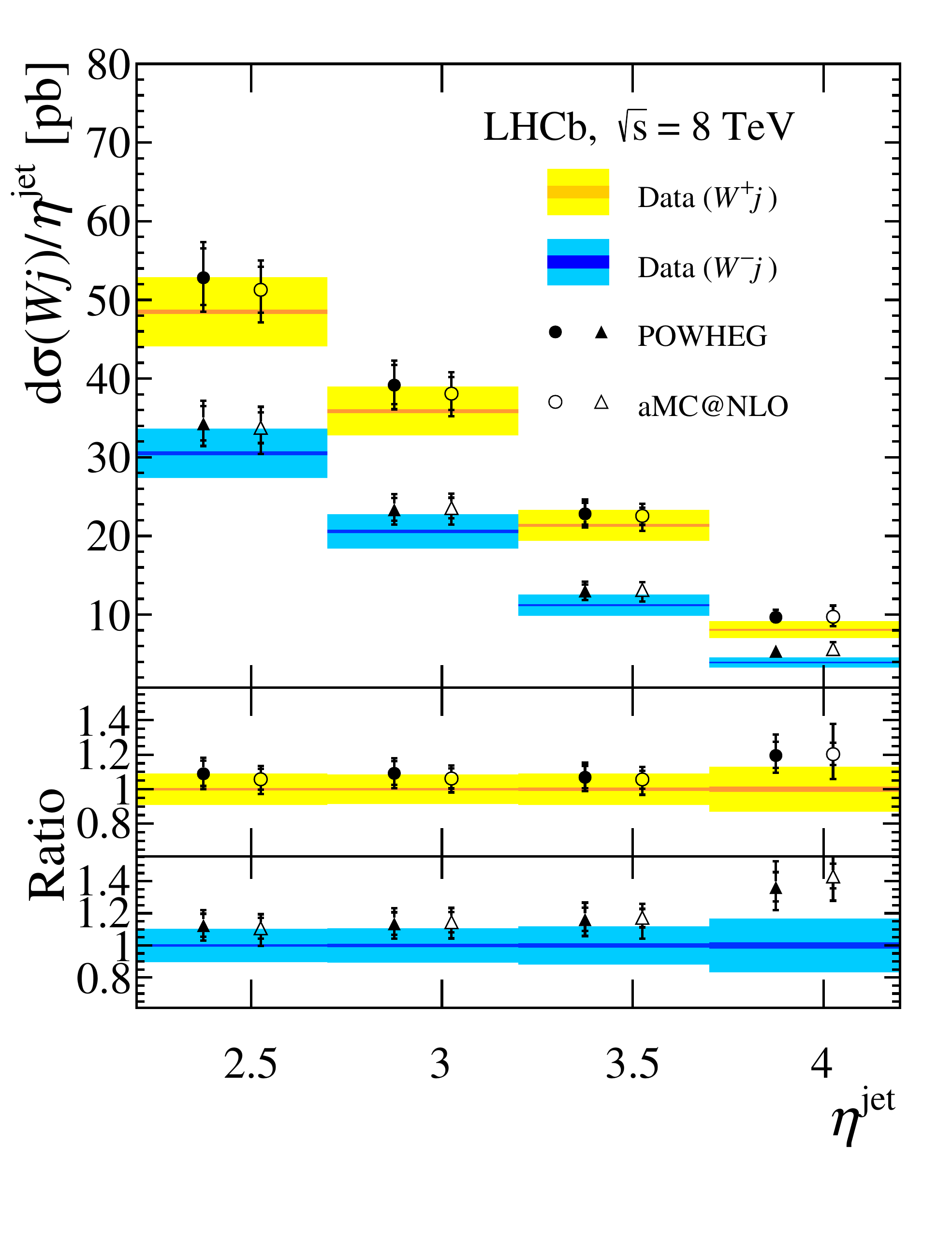}
\caption{\wj bin-averaged differential cross-sections as a function of \etamu (left) and \etaj (right). The measurements are shown as bands representing the statistical and total uncertainties, while the theoretical predictions are shown as points (displaced horizontally for presentation) representing the same bin-averaged cross-sections as the data. The inner error bar represents the scale uncertainty, and the outer error bar represents the total uncertainty. The ratio of the predicted to measured cross-sections is shown below the distribution. The \wpj and \wmj cross-sections are seen to overlap in the final bin in \etamu.}
\label{fig:wpmres_eta}
\end{center}
\end{figure}

\begin{figure}[h!]
\begin{center}
\includegraphics[width=0.495\textwidth]{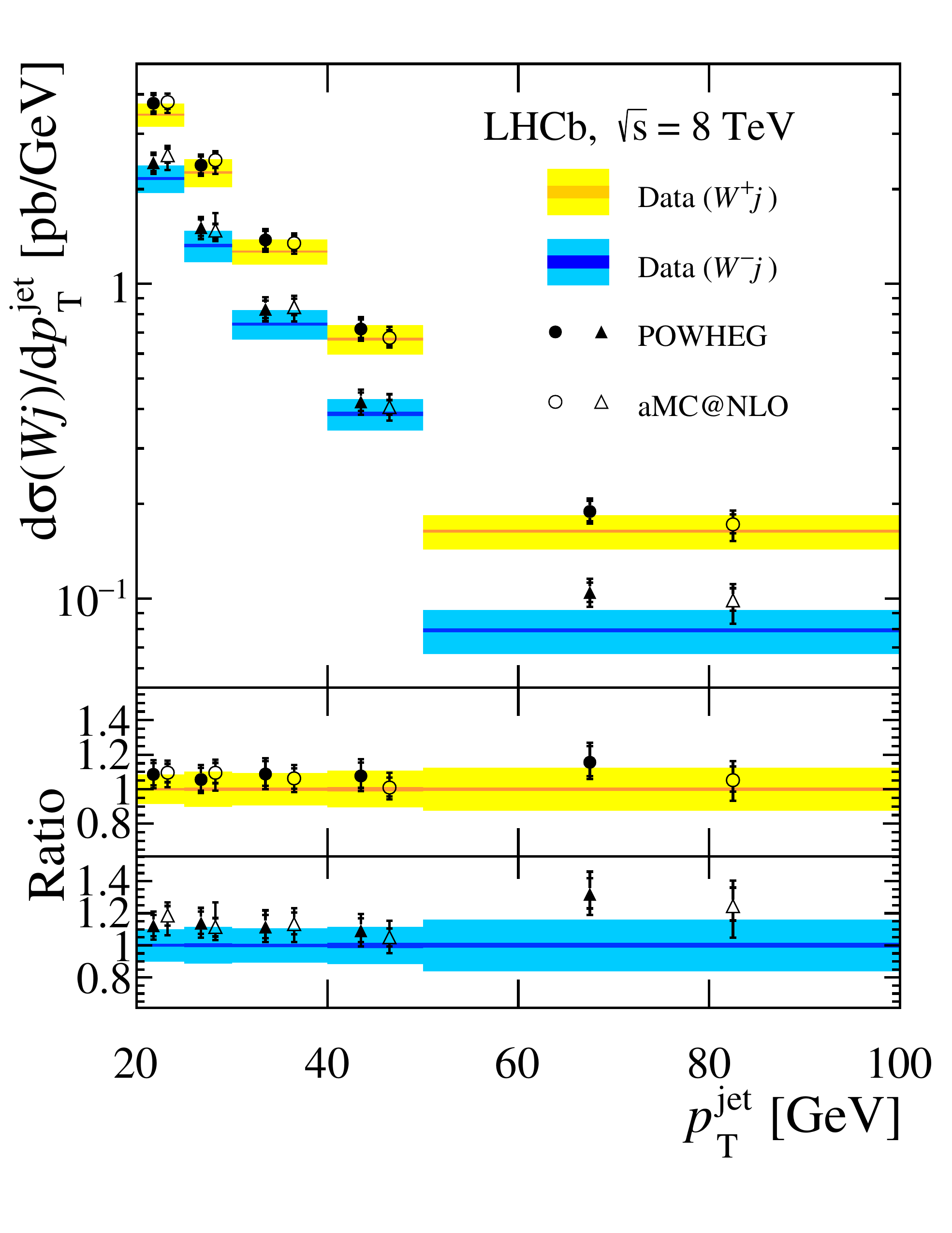}
\caption{\wj bin-averaged differential cross-sections as a function of \ptj. The experimental and theoretical components are shown as in Fig.~\ref{fig:wpmres_eta}. The ratio of the predicted to measured cross-sections is shown below the distribution.}
\label{fig:wpmres_ptj}
\end{center}
\end{figure}

\begin{figure}[ht!]
\begin{center}
\includegraphics[width=0.49\textwidth]{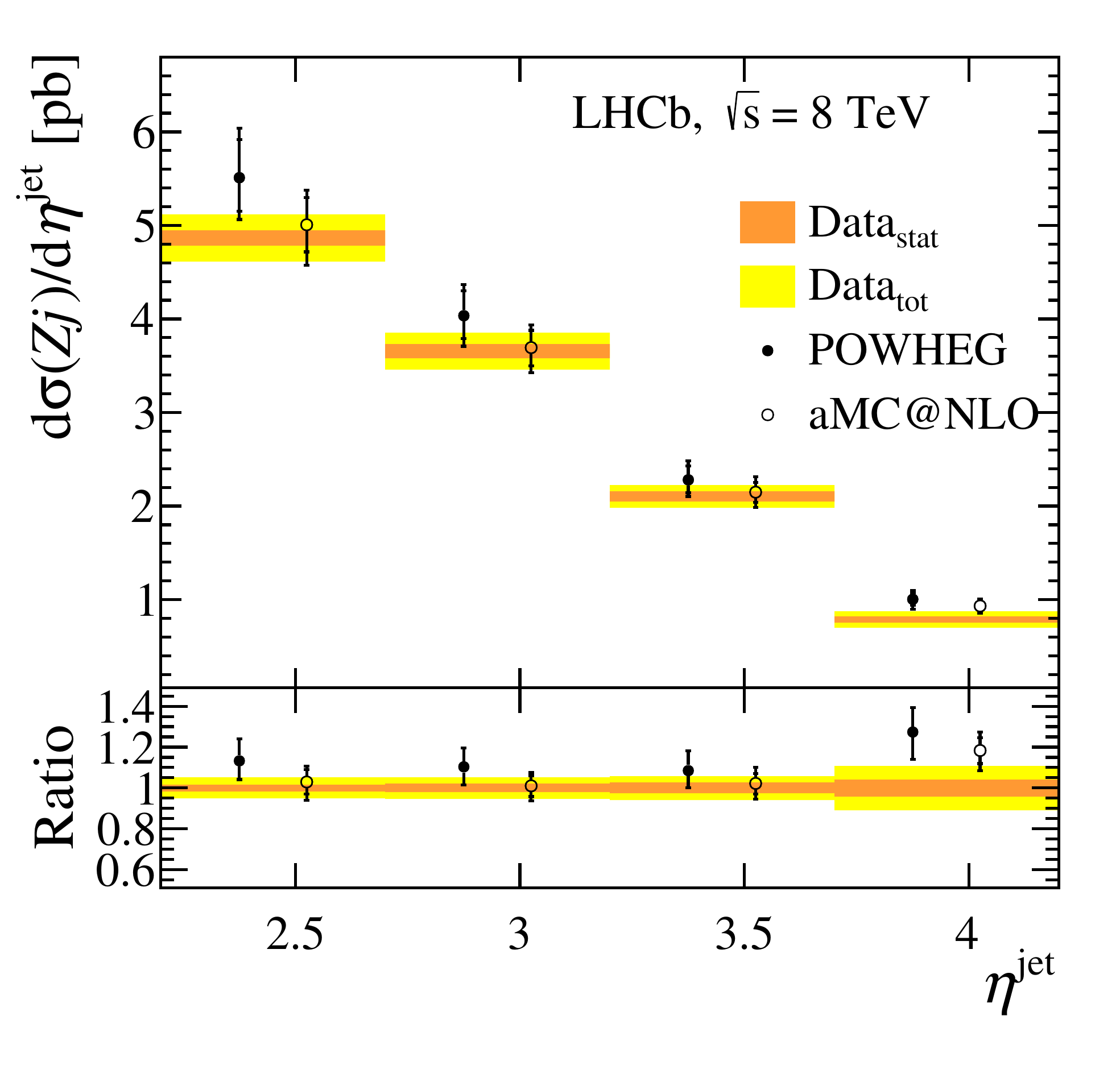}
\includegraphics[width=0.49\textwidth]{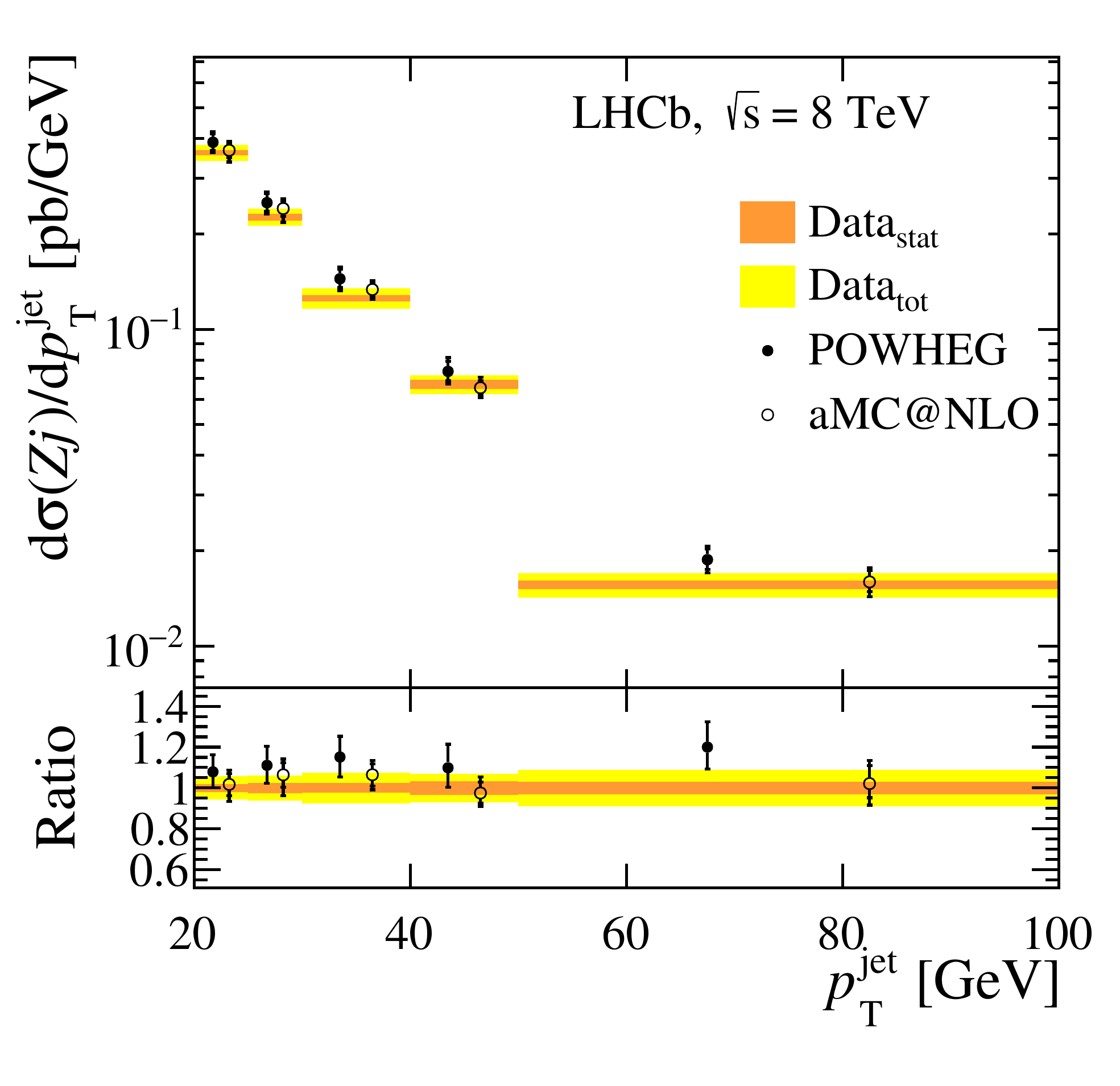}
\caption{The measured bin-averaged differential \zj production cross-section is shown as a function of  (left) \etaj and (right) \ptj. The experimental measurements are shown as bands, while the theoretical predictions are shown as points, horizontally displaced for presentation. The ratio of the predicted to measured cross-sections is shown below the distribution.}
\label{fig:zjres_1}
\end{center}
\end{figure}

\begin{figure}[t!]
\begin{center}
\includegraphics[width=0.49\textwidth]{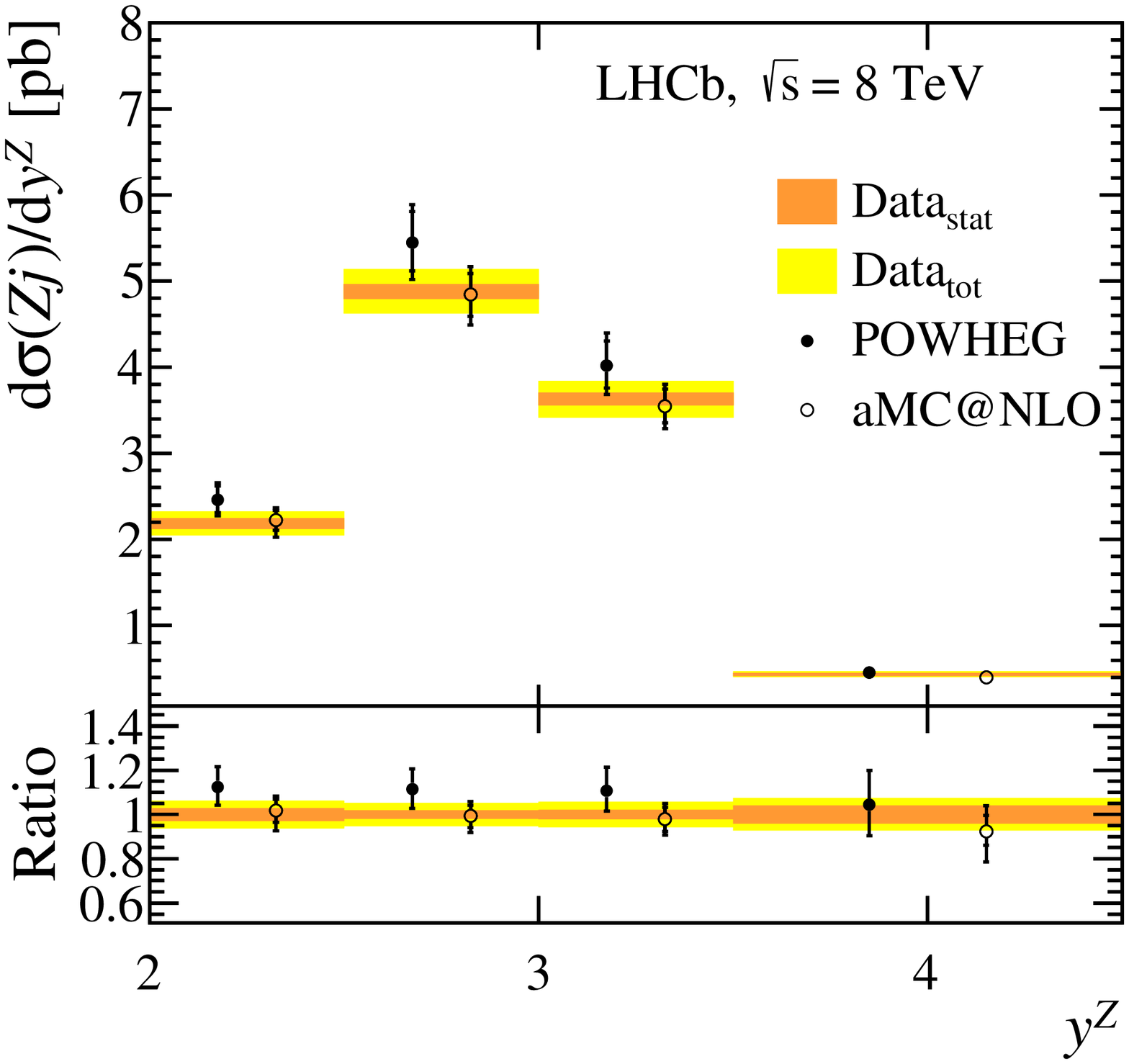}
\includegraphics[width=0.49\textwidth]{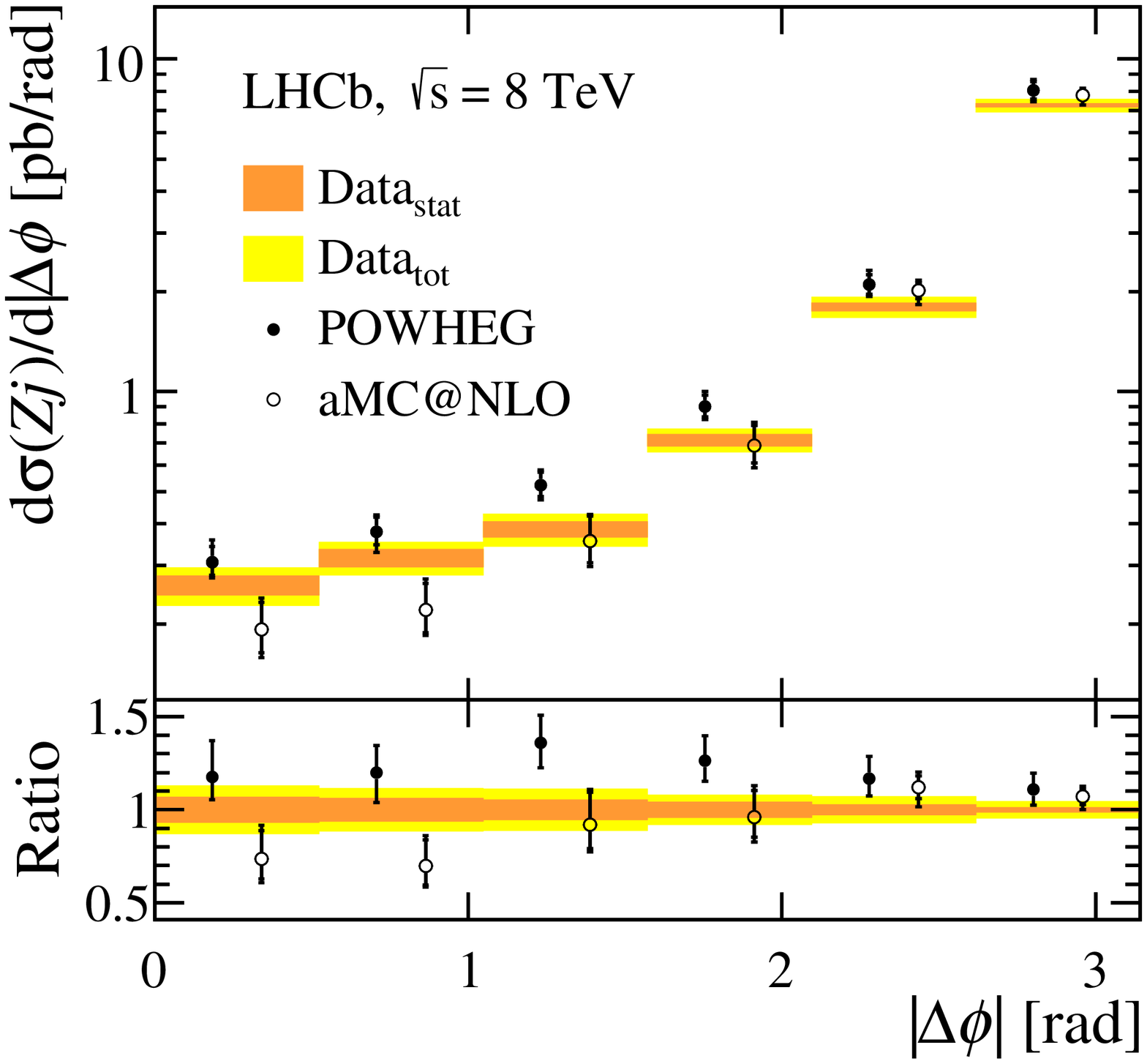}
\caption{The measured bin-averaged differential \zj production cross-section is shown as a function of  (left) \rapz and (right) azimuthal separation between the \Z boson and the jet. The experimental measurements are shown as bands, while the theoretical predictions are shown as points, horizontally displaced for presentation. The ratio of the predicted to measured cross-sections is shown below the distribution.}
\label{fig:zjres_2}
\end{center}
\end{figure}

The results are compared to theoretical predictions calculated using the \mcatnlo and \powheg generators in Fig.~\ref{fig:summary}. The uncertainty on the theoretical predictions due to higher-order effects is calculated by varying the renormalisation and factorisation scales independently by a factor of two around the nominal scale~\cite{Hamilton:2013fea}. Additional uncertainties arise from the description of the PDFs, and the value of the strong coupling, \as. The total theoretical uncertainty is obtained by combining the PDF and \as uncertainties in quadrature, and adding the result to the scale uncertainty linearly. The measurements are represented by bands where the inner band represents the statistical uncertainty and the outer band the total uncertainty. In the cross-section measurements, the scale uncertainty dominates the theoretical uncertainty, while it largely cancels in the ratios and asymmetry.
The data and predictions are further compared differentially for \wj production in Figs.~\ref{fig:wpmres_eta} and \ref{fig:wpmres_ptj}, and for \zj production in Figs.~\ref{fig:zjres_1} and \ref{fig:zjres_2}, with good agreement seen in all distributions.

\begin{figure}[ht!]
\begin{center}
\includegraphics[width=0.49\textwidth]{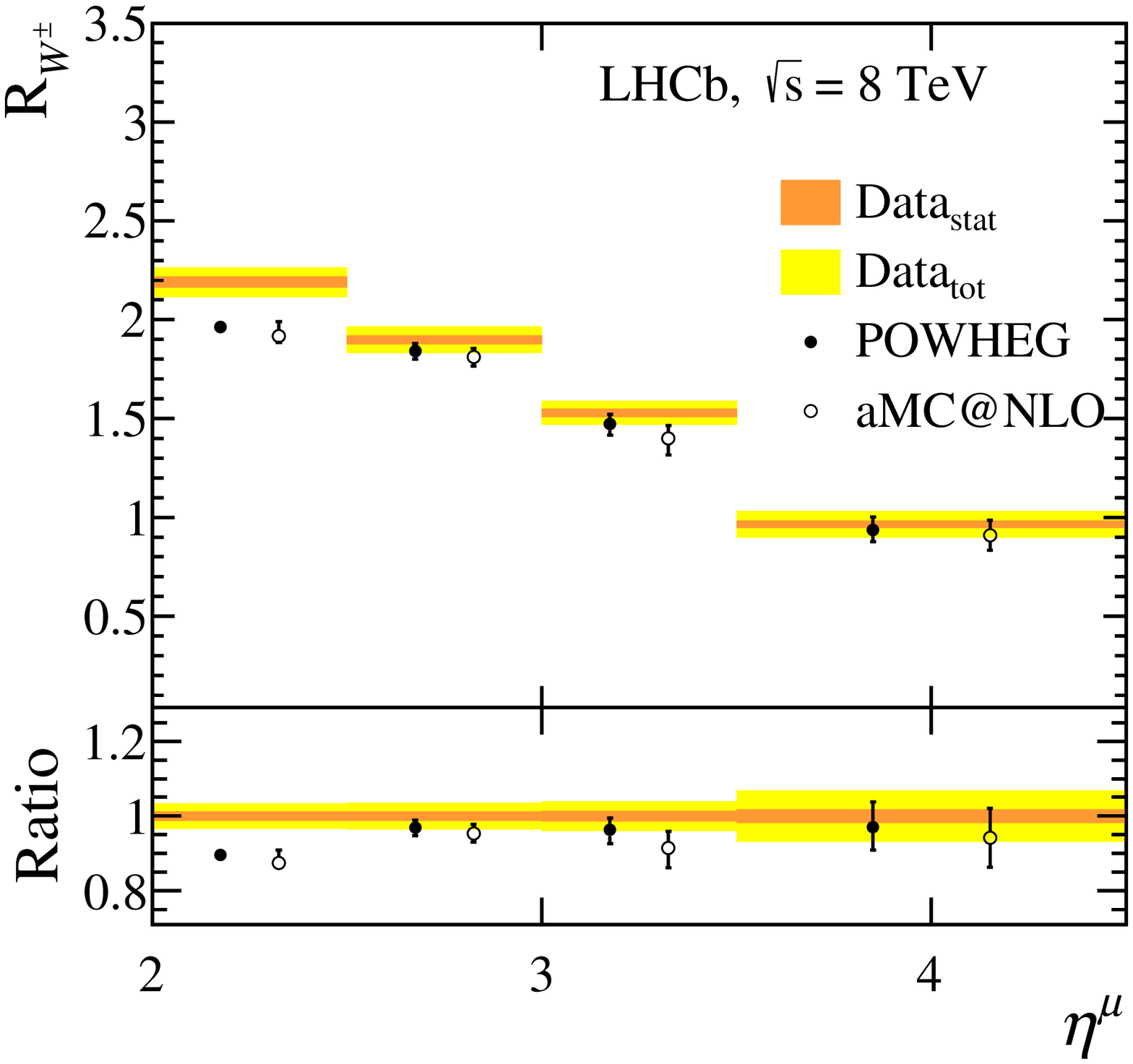}
\includegraphics[width=0.49\textwidth]{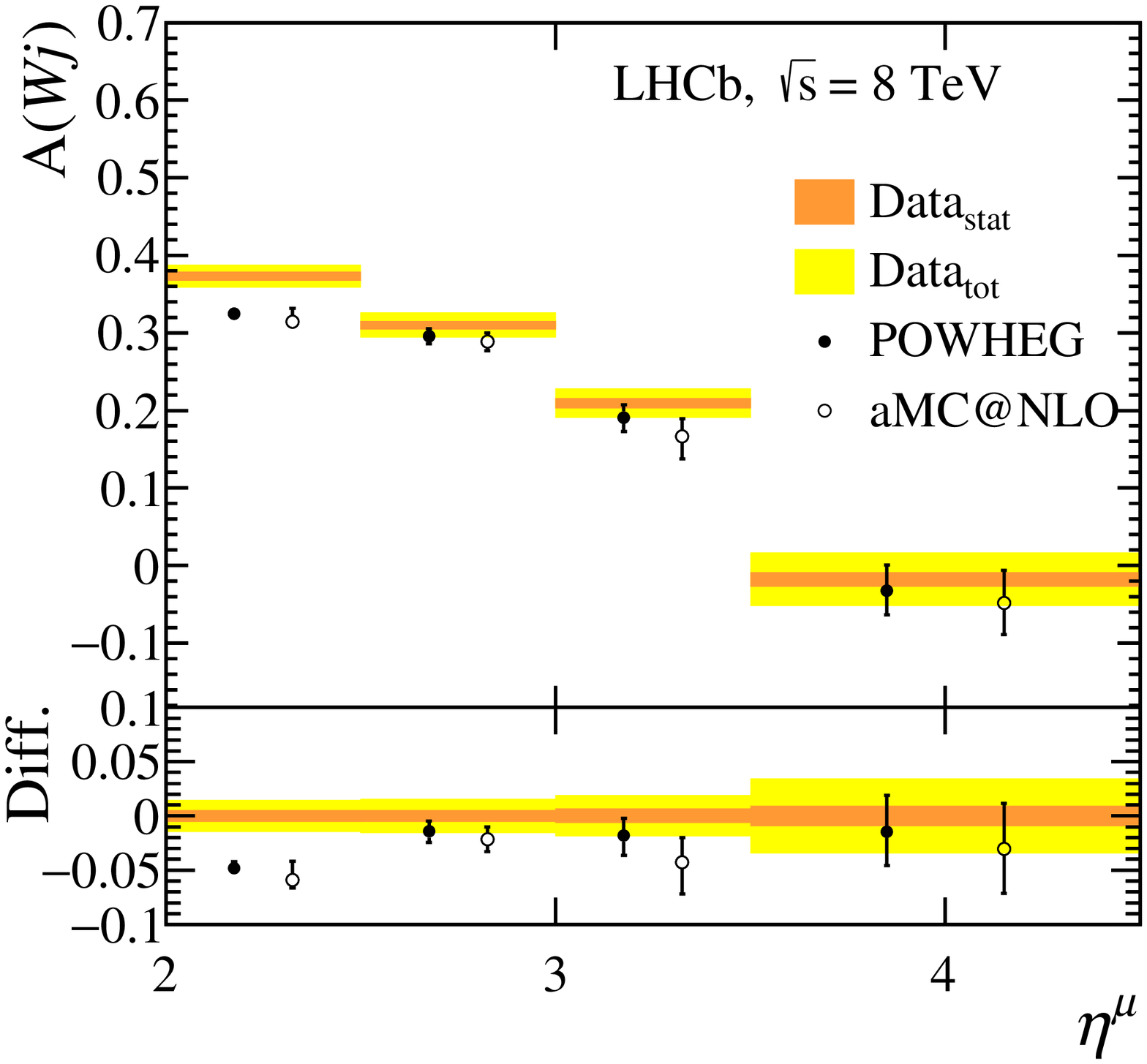}
\caption{Ratio (left) and asymmetry (right) of \wpj to \wmj production as a function of the lepton pseudorapidity. The experimental measurements are shown as bands, while the theoretical predictions are shown as points, horizontally displaced for presentation. The ratio of the predictions to the experimentally measured values is shown below the distribution for the charge ratio, while their difference is shown for the charge asymmetry.}
\label{fig:rpm_res}
\end{center}
\end{figure}

\begin{figure}[t!]
\begin{center}
\includegraphics[width=0.49\textwidth]{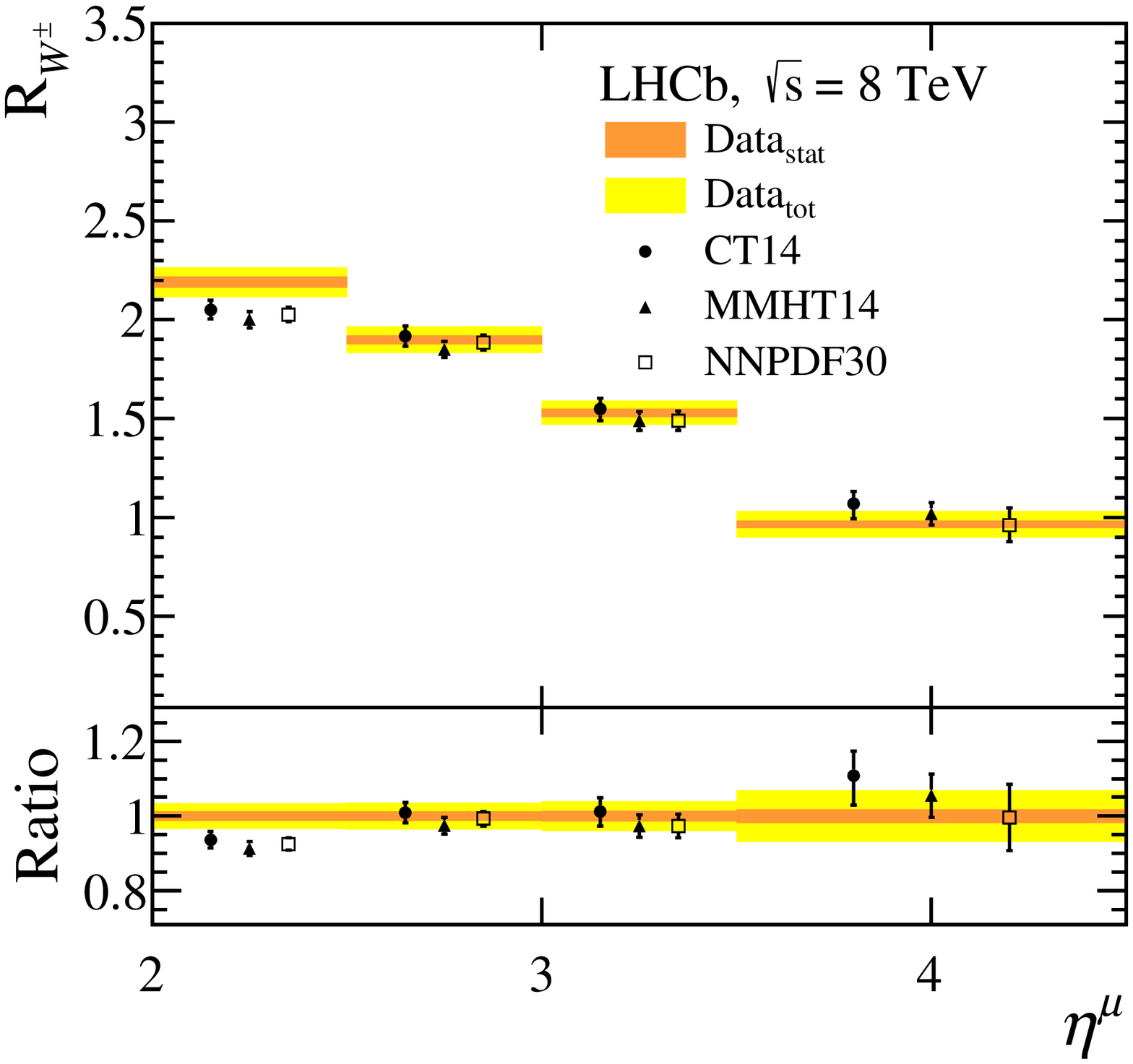}
\includegraphics[width=0.49\textwidth]{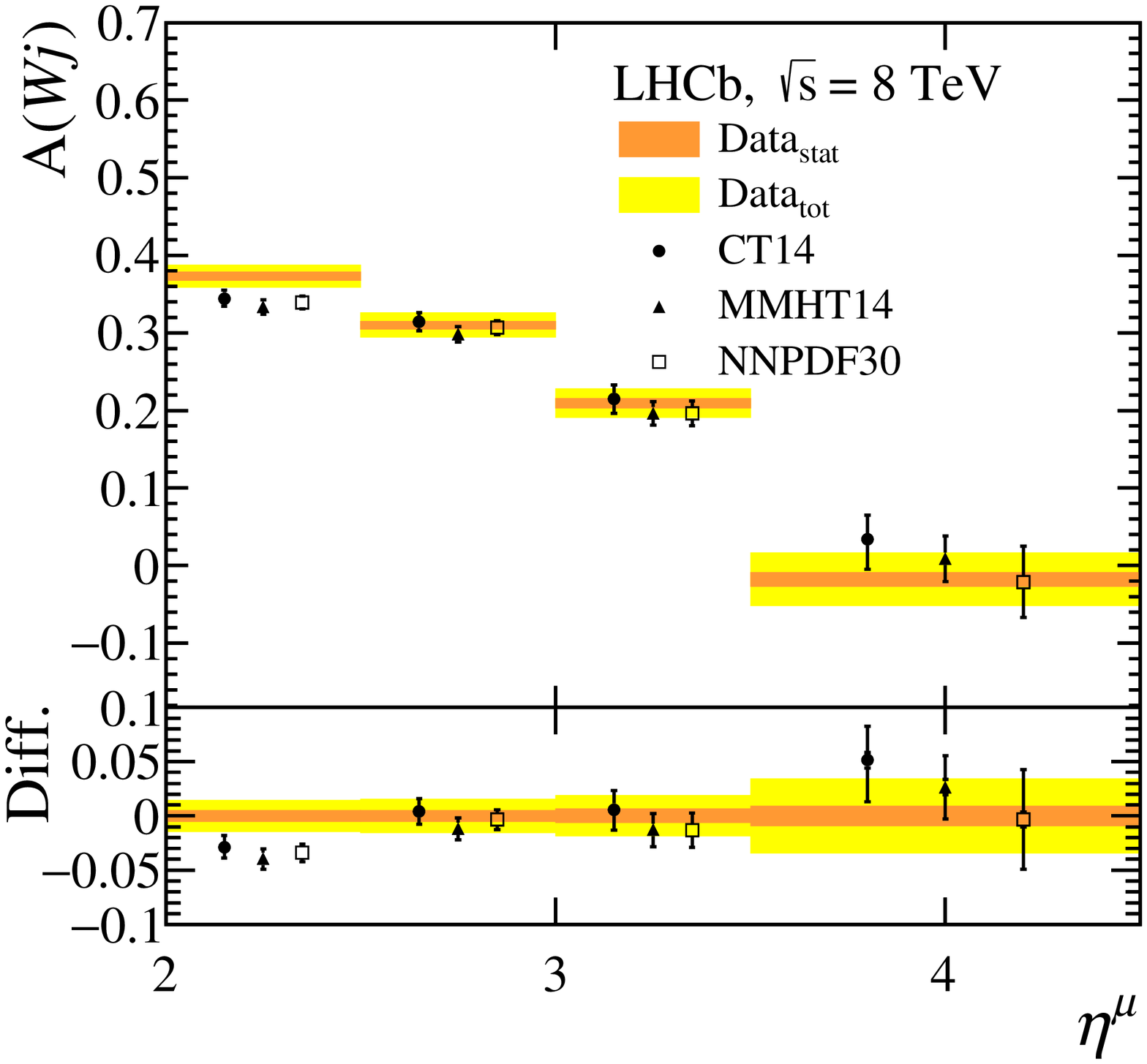}
\caption{Ratio (left) and asymmetry (right) of \wpj to \wmj production as a function of the lepton pseudorapidity compared to NLO calculations performed with the \fewz generator and three different PDF sets. The experimental measurements are shown as bands, while the theoretical predictions are shown as points, horizontally displaced for presentation. The same comparisons are shown below the distribution as described in Fig.~\ref{fig:rpm_res}.}
\label{fig:rpm_fewz}
\end{center}
\end{figure}

Further to the total and differential production cross-sections, measurements of the charge ratio and asymmetry of \wj production are also performed as a function of lepton pseudorapidity and are compared to \powheg and \mcatnlo in Fig.~\ref{fig:rpm_res}. Due to the cancellation of scale uncertainties, these distributions are expected to show sensitivity to the PDFs and consequently are also compared in Fig.~\ref{fig:rpm_fewz} to fixed-order calculations performed with \fewz separately for the NNPDF3.0, CT14 and MMHT14 PDF sets. The fixed-order predictions are expected to give a good description of the ratios and asymmetries as the effects of higher-order terms and hadronisation largely cancel between the positively and negatively charged \wj predictions. In general, good agreement is seen between the data and the predictions, although the data presents a slightly larger ratio and asymmetry, particularly in the first bin of \etamu. However, when the spread of predictions obtained using different PDF sets is considered, the deviations are not significant.

\clearpage

%%%%%%%%%%%%%%%%%%%%%%%%%%%%%%%%%%%%
% !TEX root = main.tex
%%%%%%%%%%%%%%%%%%%%%%%%%%%%%%%%%%%%
\section{Conclusions}

Measurements of the forward \W and \Z boson cross-sections in association with jets at \mbox{\sqs = 8\tev} are presented. The \W bosons are reconstructed in the decay $\W\to\mu\nu_\mu$ and the \Z bosons in the decay $\Z\to\mu\mu$.
Total cross-sections are presented in the forward fiducial region in addition to measurements of the charge ratio and asymmetry of \wj production and the ratio of \wj to \zj production. Differential cross-sections are presented as a function of \ptj, \etaj, \etamu in the case of \wj production, and for \zj production, where a full reconstruction of the final state is possible, measurements are presented as a function of \ptj, \etaj, \rapz, and the azimuthal separation of the \Z boson and the jet, \dphi. The \wj charge ratio and asymmetry are presented as a function of \etamu. All measurements are observed to be in agreement with predictions obtained at $\mathcal{O}(\alpha_s^2)$ interfaced with a parton shower in order to achieve NLO plus leading-log accuracy. The measurements of the charge ratio and asymmetry of \wj production are also compared to predictions obtained at $\mathcal{O}(\alpha_s^2)$ in fixed order perturbative QCD and show good agreement.

\clearpage

% Do not include this in analysis note and conference reports
\section*{Acknowledgements}

\noindent We express our gratitude to our colleagues in the CERN
accelerator departments for the excellent performance of the LHC. We
thank the technical and administrative staff at the LHCb
institutes. We acknowledge support from CERN and from the national
agencies: CAPES, CNPq, FAPERJ and FINEP (Brazil); NSFC (China);
CNRS/IN2P3 (France); BMBF, DFG and MPG (Germany); INFN (Italy); 
FOM and NWO (The Netherlands); MNiSW and NCN (Poland); MEN/IFA (Romania); 
MinES and FANO (Russia); MinECo (Spain); SNSF and SER (Switzerland); 
NASU (Ukraine); STFC (United Kingdom); NSF (USA).
We acknowledge the computing resources that are provided by CERN, IN2P3 (France), KIT and DESY (Germany), INFN (Italy), SURF (The Netherlands), PIC (Spain), GridPP (United Kingdom), RRCKI and Yandex LLC (Russia), CSCS (Switzerland), IFIN-HH (Romania), CBPF (Brazil), PL-GRID (Poland) and OSC (USA). We are indebted to the communities behind the multiple open 
source software packages on which we depend.
Individual groups or members have received support from AvH Foundation (Germany),
EPLANET, Marie Sk\l{}odowska-Curie Actions and ERC (European Union), 
Conseil G\'{e}n\'{e}ral de Haute-Savoie, Labex ENIGMASS and OCEVU, 
R\'{e}gion Auvergne (France), RFBR and Yandex LLC (Russia), GVA, XuntaGal and GENCAT (Spain), Herchel Smith Fund, The Royal Society, Royal Commission for the Exhibition of 1851 and the Leverhulme Trust (United Kingdom).

%\input{appendix}

%\input{supplementary-app}

% This should be taken out in the final paper

\addcontentsline{toc}{section}{References}
\setboolean{inbibliography}{true}
\bibliographystyle{LHCb}
\bibliography{main,LHCb-PAPER,LHCb-CONF,LHCb-DP,LHCb-TDR}

\newpage

% Author List ----------------------------                                                                                                                                                                                                                                                                                                
%  You need to get a new author list!                                                                                                                                                                                                                                                                                                    

%\input{LHCb_HD_authorlist_2014-06-20}
 
\newpage
\centerline{\large\bf LHCb collaboration}
\begin{flushleft}
\small
R.~Aaij$^{39}$,
C.~Abell{\'a}n~Beteta$^{41}$,
B.~Adeva$^{38}$,
M.~Adinolfi$^{47}$,
Z.~Ajaltouni$^{5}$,
S.~Akar$^{6}$,
J.~Albrecht$^{10}$,
F.~Alessio$^{39}$,
M.~Alexander$^{52}$,
S.~Ali$^{42}$
% ...

{\footnotesize \it
% ...
$ ^{5}$Clermont Universit{\'e}, Universit{\'e} Blaise Pascal, CNRS/IN2P3, LPC, Clermont-Ferrand, France\\
$ ^{6}$CPPM, Aix-Marseille Universit{\'e}, CNRS/IN2P3, Marseille, France\\
% ...
$ ^{10}$Fakult{\"a}t Physik, Technische Universit{\"a}t Dortmund, Dortmund, Germany\\
% ...
$ ^{38}$Universidad de Santiago de Compostela, Santiago de Compostela, Spain\\
$ ^{39}$European Organization for Nuclear Research (CERN), Geneva, Switzerland\\
% ...
$ ^{41}$Physik-Institut, Universit{\"a}t Z{\"u}rich, Z{\"u}rich, Switzerland\\
$ ^{42}$Nikhef National Institute for Subatomic Physics, Amsterdam, The Netherlands\\
% ...
$ ^{47}$H.H. Wills Physics Laboratory, University of Bristol, Bristol, United Kingdom\\
% ...
$ ^{52}$School of Physics and Astronomy, University of Glasgow, Glasgow, United Kingdom\\
% ...
}
\end{flushleft}

\end{document}